\documentclass[useAMS,usenatbib]{mn2e}
\usepackage{graphicx}

\newcommand {\ha}    {\ifmmode ${H}$\alpha \else {H}$\alpha$\fi}
\newcommand {\hb}    {\ifmmode ${H}$\beta  \else {H}$\beta$\fi}
\newcommand {\hg}    {\ifmmode ${H}$\gamma \else {H}$\gamma$\fi}
\newcommand {\laa}   {\ifmmode ${L}$\alpha \else {L}$\alpha$\fi}
\newcommand {\ergs}  {\ifmmode ${\ergs}$ \else ergs\,cm$^{-2}$\,s$^{-1}$\fi}
\newcommand {\ergsA} {\ergs\,\ifmmode ${\AA}$^{-1} \else \AA$^{-1}$\fi}
\newcommand {\MSun}  {\ifmmode ${M}$_\Sun \else {M}$_\Sun$\fi}
\newcommand {\degC}  {$^{\rm o}$C}

\title[Black Hole Mass in Mrk 6]{BLR kinematics and Black Hole Mass in Markarian 6}
\author[V.~T.~Doroshenko et al.]{V.~T.~Doroshenko,$^{1,2,3}$\thanks{E-mail: \mbox{vdorosh@sai.crimea.ua (VTD);}\hfill \mbox{~} sergeev@crao.crimea.ua (SGS).}
S. G. Sergeev,$^{1,3}$\footnotemark[1] S.~A.~Klimanov,$^{4}$
V.~I.~Pronik,$^{1,3}$
\newauthor Yu.~S.~Efimov$^{1}$\thanks{Deceased, 2011 October 21.}\\
$^{1}$Crimean Astrophysical Observatory, P/O Nauchny Crimea 98409, Ukraine\\
$^{2}$Crimean Laboratory of the Sternberg Astronomical Institute, P/O Nauchny, 98409 Crimea, Ukraine\\
$^{3}$Isaak Newton Institute of Chile, Crimean Branch, Ukraine\\
$^{4}$Central Astronomical Observatory of Pulkovo, Pulkovskoe Shosse 65, 196140 St. Petersburg, Russia
}

\begin{document}
\date{Accepted . Received  ; in original form }
\pagerange{\pageref{firstpage}--\pageref{lastpage}} \pubyear{2012}
\maketitle
\label{firstpage}

\begin{abstract}

We present results of the optical spectral and photometric observations of the nucleus of Markarian~6 made with the 2.6-m Shajn telescope at the Crimean Astrophysical Observatory. The continuum and emission Balmer line intensities varied more than by a factor of two during 1992--2008. The lag between the continuum and \hb\ emission line flux variations is $21.1 \pm 1.9$ days. For the \ha\ line the lag is about 27 days but its uncertainty is much larger. We use Monte-Carlo simulation of the random time series to check the effect of our data sampling on the lag uncertainties and we compare our simulation results with those obtained by random subset selection (RSS) method of \citet{Peterson98}. The lag in the high-velocity wings are shorter than in the line core in accordance with the virial motions. However, the lag is slightly larger in the blue wing than in the red wing. This is a signature of the infall gas motion. Probably the BLR kinematic in the Mrk~6 nucleus is a combination of the Keplerian and infall motions. The velocity-delay dependence is similar for individual observational seasons.  The measurements of the \hb\ line width in combination with the reverberation lag permits us to determine the black hole mass, $M_{BH}=(1.8\pm0.2)\times 10^8 M_{\sun}$. This result is consistent with the AGN scaling relationships between the BLR radius and the optical continuum luminosity ($R_{BLR}\propto L^{0.5}$) as well as with the black-hole mass--luminosity relationship (M$_{BH}-L$) under the Eddington luminosity ratio for Mrk~6 to be $L_{bol}/L_{Edd}\sim 0.01$.
\end{abstract}

\begin{keywords}
galaxies: active -- galaxies: nuclei -- galaxies: Seyfert -- galaxies: individual: Mrk~6
\end{keywords}

\section[]{Introduction}

Over the past nearly 30 years the method of reverberation mapping (RM) (\citealt{Peterson88} and reference therein) has become one of the standard methods for studying the Active Galactic Nuclei (AGNs). It is based on the assumption that in a typical Seyfert galaxy the source of continuous radiation near a black hole, named as accretion disc (AD), is expected to be of order $10^{13}-10^{14}$~cm. Photoionization of the gas located at a distance of order $10^{16}$~cm produces broad emission lines. The relationship between the continuum and emission line fluxes can be represented by the equation \citep{BMcKee}
\[
L(t)=\int \Psi(\tau) C(t-\tau) dt,
\]
where $C(t)$ and $L(t)$ are the observed continuum and emission-line light curves,
and $\Psi(\tau)$ is the 1-d transfer function (TF). The TF determines the emission line
response to a $\delta$-function continuum pulse as seen by a distant observer. So, the emission lines "echo" or "reverberate" in response to the continuum changes with a delay $\tau$. The size
of the region where broad lines (BLR) are formed can be written as $R_{BLR}=c\tau$. The primary
task of the RM method is to use the observable $C(t)$ and $L(t)$ to solve the above integral
equation for the TF in order to obtain information about the geometry and physical conditions
in the BLR. Unfortunately, it is very difficult to find a unique and reliable solution to this
equation. However, it is possible to find a temporal shift (lag) between the continuum and
emission line light curves using the cross-correlation analysis.

Applying the virial assumption, the mass of the black hole can be determined when the BLR size
and the velocity dispersion of the BLR gas are known \citep{Peterson04}. The present tremendous
progress being made in black hole mass estimates can be attributed to the reverberation method.

On the other hand, different segments of a single emission line seem to be formed at
different effective distances from the ionizing source. In that case, the response in the flux
of emission line at line-of-sight velocity $V_r$ and time delay $\tau$ is caused by the 2-d
transfer function or "velocity delay map" \citep{Horne04}. The reverberation technique applied
to different parts of a single emission line allows one to make conclusions about the velocity field
of the BLR gas. To the present time, considerable progress was made in understanding the
direction of the BLR gas motion \citep{Gaskell09, Bentz08, Denney09, Bentz10}. For example,
some distinctive signatures for infalling gas motions in NGC~3516 and Arp~151 \citep{Bentz09b, Denney09} were revealed: the blue side of the line lagging the red side. NGC~5548 shows the virialized gas motions with the symmetric lags on both the red and blue sides of line. However, the BLR gas in NGC~3227 shows  the signature of radial outflow: shorter lags for the blue-shifted gas and longer lags for the red-shifted gas \citep{Denney09}.

More than 40 AGNs have been studied by the RM up to now \citep{Peterson04, Bentz09b, Denney10}.
However, the Mrk~6 nucleus is absent in this list. Just a few studies of this galaxy have been
made by the reverberation method. We can mention the paper by \citet{DS03} based on the archive
spectra of Mrk~6 obtained from 1970--1991 using the image tube spectrograph at the 2.6-m
telescope of the Crimean Astrophysical Observatory (CrAO). There is also another paper by
\citet{Ser99} that includes the results of the 1992--1997 observations with the same spectrograph but with a CCD detector. In this paper a lag in the flux variations of hydrogen lines with respect to the adjacent continuum flux variations was reported for the first time and changes in the line profiles were studied.

Mrk~6 is a Seyfert 1.5 galaxy (Sy~1.5). This galaxy was one of the first galaxies in which the
strong variability of the \hb\, emission line profile was detected by \citet{KhW71}. Although
there have not been many optical observations of Mrk~6, there are some radio studies of Mrk~6
available \citep{Kuk96, Kharb06}, which revealed the complex structure of the radio emission. The
X-ray emission from the Mrk~6 nucleus \citep{Feldmeier99, Malizia03, Immler03, Schurch06}
exhibits a complex X-ray absorption and some authors assume that the BLR is a possible location
of this absorption complex \citep{Malizia03}.

In this paper we present the results of our optical spectroscopic observations of Mrk~6 during
the monitoring campaign from 1998 to 2008 performed after publishing our earlier papers with
observations made from 1970--1997. For completeness we apply our analysis to all the spectral
CCD observations that have been made since 1992 and we also use the results of our photometric
observations in the $V$ band. Throughout the entire work, we take $z=0.01865$ from our spectral estimates, the distance to Mrk~6 equal to $D=81$~Mpc, and $H_0=$70~km\,s$^{-1}~Mpc^{-1}$. The
observations and data reduction are described in Section~2. We present the cross-correlation
analysis in Section~\ref{fccf}, the line width measurements in Section~\ref{linewidth}; the estimates of the black hole mass, the mass--luminosity and lag--luminosity diagrams are presented in Sections~\ref{mass-bh} and \ref{slm}, the velocity-resolved reverberation lag analysis is performed in Section~\ref{vrrlag}. The results are summarized in Section~\ref{sum}.

\section{Observations}
\subsection{Spectral observations and data processing}
The \ha\ and \hb\ spectra of the Seyfert galaxy Mrk~6 were obtained from the CrAO 2.6-m Shajn
telescope. Prior to 2005 we used the Astro-550 CCD which had a size 580$\times520$~pixels and was
cooled by liquid nitrogen. The dispersion was 2.2~\AA\,pixel$^{-1}$, and the spectral resolution
was about 7--8~\AA. The working wavelength range was about 1200~\AA. The entrance slit width was
3\arcsec. For technical reasons it was not always possible to set the same position angle (PA) of the
entrance slit. Most of the observations were performed along PA$\sim125^{\rm o}$ or
PA$\sim90^{\rm o}$. The typical exposure time was about one hour. The "extraction window" was
equal to 11\arcsec.

In July 2005 the Astro-550 CCD was replaced with the SPEC-10 $1340\times100$~pixel CCD,
thermoelectrically cooled up to $-100$\degC. In this case the dispersion was
1.8~\AA\,pixel$^{-1}$. A 3\farcs0 slit with a 90\degr position angle was utilized for these
observations. The higher quantum efficiency (95\% max.) and the lower read noise of this CCD
allowed us to obtain higher quality spectra under shorter exposure times. The spectral wavelength
range for these data sets was about 2000~\AA\, near the \ha\, and \hb\ regions. However, the red
and blue edges of the CCD frame are unusable because of vignetting. Generally, all spectra in
both spectral regions were obtained with a single exposure during the whole night. Our final
data set for 1998--2008 consists of 135 spectra in the \hb\, region and 48 spectra in \ha.
The mean signal-to-noise ratio (S/N) is equal to 32 for the Astro-550 and 73 for the SPEC-10 in
the \hb\, region and is equal to 40 and 66 in the \ha\, region, respectively.

As a rule, each observation was preceded by four short-time ($\sim$10~s) exposures of the
standard star BS~3082, whose spectra were obtained at approximately the same zenith distance as
that of the galaxy. The spectral energy distribution for these spectra was taken from
\citet{Khar88}. The standard star spectra were used to remove telluric absorption features from
the spectra of Mrk~6, to provide the relative flux calibration, and to measure the seeing
parameter defined as the FWHM of the cross-dispersion profile on the CCD image. The description
of the primary data processing as well as some details of absolute calibration and measurements
of the spectra are given in \citet{Ser99}.

\begin{figure}
\includegraphics[width=84mm]{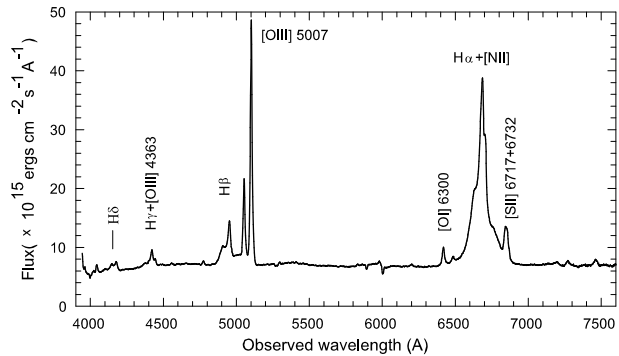}
 \caption{Mean Mrk~6 spectrum obtained by combining the quasi-simultaneous pairs of spectra
from the \ha\, and \hb\, spectral regions (SPEC-10 CCD only).}
 \label{M6-spe}
\end{figure}

\begin{figure}
\begin{center}
\includegraphics[width=60mm]{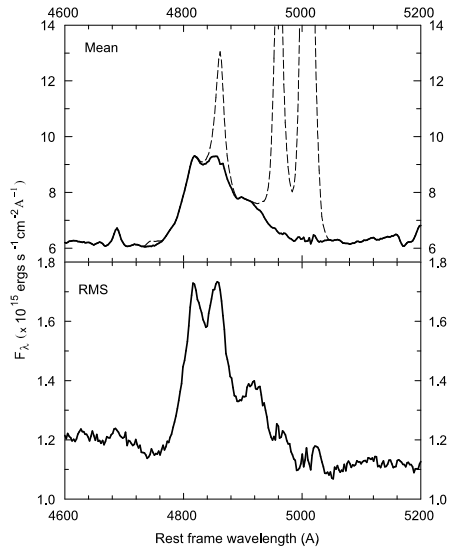}
 \caption{Mean and rms spectra of Mrk~6 in the \hb\ region from our observations 1992--2008.}
 \label{M6mn-rms}
 \end{center}
\end{figure}

The flux calibration was carried out by assuming the narrow emission line fluxes to be constant.
We chose the narrow [OIII]$\lambda$5007 (for the \hb\, region) and [SII]$\lambda\lambda$6717,~6732 emission lines (for the \ha\, region) as the internal flux
standards. Their absolute fluxes were measured from the spectra obtained under photometric
conditions by using the spectra of the comparison star BS~3082. The mean fluxes in these lines
are given in \citet{Ser99}: F([OIII]$\lambda$5007)=$(6.90\pm0.11)\times10^{-13}$~erg~s$^{-1}$~cm$^{-2}$;
F[SII]~$\lambda\lambda(6717+6731)$=(1.61$\pm0.09)\times10^{-13}$~erg~s$^{-1}$~cm$^{-2}$, and
F[OI]~$\lambda$6300=(0.597$\pm0.035)\times10^{-13}$~erg s$^{-1}$~cm$^{-2}$. The [SII] lines
reside on the far wings of the broad \ha\, line. Thus, to measure their fluxes, we selected the
pseudo-continuum zones closely spaced around each line. The line fluxes were measured by
integrating the spectra over the specified wavelength intervals and above the continuum (or
local pseudo-continuum), which was fitted with a straight line in the selected zones. The mean
continuum flux per unit wavelength was determined in two windows: at 5162--5186~\AA\,
(designated as $F5170$) and 6985--7069~\AA\, (designated as $F7030$). The continuum zones and
integration limits are the same as in \citet{Ser99}. The line and continuum flux uncertainties contain errors related to the S/N of the source spectra, atmospheric dispersion, changes in the position angle of the slit, and seeing effects. Evaluation of these uncertainties is considered
in \citet{Ser99}. The mean spectrum of Mrk~6 produced by combining 23 quasi-simultaneous pairs of
spectra from the \ha\ and \hb\ regions is shown in Fig.\ref{M6-spe}. Figure \ref{M6mn-rms} shows the mean and rms spectra of Mrk~6 based on our observations in 1992--2008.

\subsection{Optical photometry}
\label{oph} In order to improve the time resolution of our data set in the F5170 continuum we
added the $V$-band photometry to our spectral observations. Photometric data came from two
sources: the $UBV$ observations were made at the Crimean Laboratory of the Sternberg
Astronomical Institute of Moscow University, and the $BVRI$ observations were obtained at the
CrAO. The $UBV$ observations were obtained in the standard Johnson photometric system and were carried out from 1986 to 2009 at the 60-cm Zeiss telescope with a photo-multiplier detector through the aperture A=27$\farcs5$. The mean uncertainty of Mrk~6 in the $V$-band is 0\fm022. These data were partially published by \citet{Dor2003}.

In 2001 we started regular observations of Mrk~6 using the CrAO 70-cm AZT-8 telescope and the
AP7p CCD. The CCD field covers 15\arcmin$\times$15\arcmin. Photometric fluxes were measured
within an aperture of 15\farcs0. The mean uncertainty of the $V$-band CCD observations is
0\fm009. Further details about the instrumentation, reductions, and measurements of the $BVRI$
photometric data can be found in \citet{dor05}. These data were partially published by
\citet{Ser05}.

\subsection{Light curves}
\label{comlc} 
\begin{figure}
\includegraphics[width=84mm]{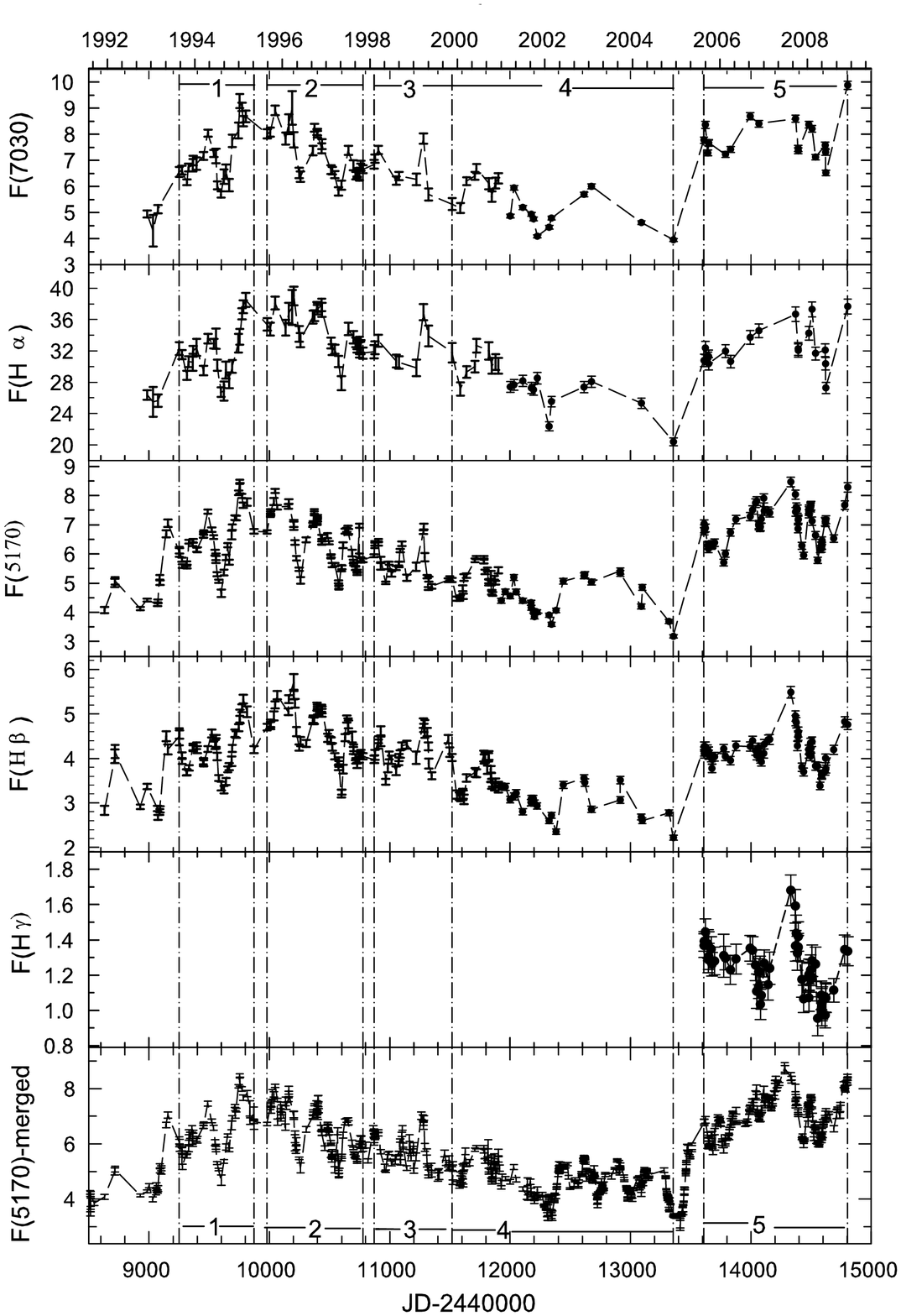}
\caption{Light curves of Mrk~6 shown for the continuums and \ha, \hb, and \hg\ lines.
Narrow lines were not subtracted from the fluxes. Bottom panel shows the merged continuum light curve used for cross-correlation analysis. Units are 10$^{-13}$~erg~cm$^{-2}$~s$^{-1}$ and
10$^{-15}$~erg~cm$^{-2}$~s$^{-1}$~\AA$^{-1}$ for the lines and continua, respectively. 
The vertical dash-dotted lines show the boundaries of the five time intervals considered in the present paper.}
\label{lc}
\end{figure}

Light curves in \ha\, and \hb, and the adjacent continuum are shown in Fig.~\ref{lc} for spectral observations corrected for seeing.
The continuum light curves obtained from the photometric $V$-band observations were scaled to
the flux density measured from the spectroscopic observations. To this end, we used the
observations made on the same nights almost simultaneously. We have 29 appropriate observational
nights at the 2.6-m and 70-cm telescopes (spectra plus CCD photometry) and 40 appropriate nights
at the 2.6-m and 60-cm telescopes (spectra plus the $UBV$ photoelectric photometry). The
correlation coefficient between the spectral continuum and the $V$-band CCD flux for the
appropriate nights is $r$=0.984 (n=29 points), and the correlation coefficient between the
spectral continuum and the $V$-band flux from the $UBV$ observations is $r$=0.993 (n=40 points).
Using the regression equations, we converted our $V$-band photometric fluxes to the spectral
continuum fluxes (F5170). For the nights where both spectral and photometric observations were
available, the continuum fluxes are calculated as the weighted average.  The fluxes were not
corrected for the host starlight contamination and Galactic reddening.

{\bf Tables~\ref{hb-hgflx} and \ref{haflx} give the light curves from the spectral observations for 1998--2008 \citep[the data for 1991--1997 are available in][]{Ser99}.
The \hb\ and \hg\ line fluxes together with the spectral continuum fluxes $F_{5170}$
are shown in Table~\ref{hb-hgflx}. The \ha\ line fluxes and the spectral continuum fluxes $F_{7030}$
are shown in Table~\ref{haflx}. Table~\ref{cntflx} gives combined continuum fluxes from both the spectral observations and from $V$-band photometric measurements for 1991--2008. All the fluxes are seeing-corrected. These light curves have been used for the subsequent time-series analysis.}

\begin{table}
  \caption{F5170 spectral continuum, the H$\beta$ and H$\gamma$ line fluxes.}
  \label{hb-hgflx}
  \begin{tabular}{@{}cccc}
\hline
JD-2,440,000  & $F5170$   & H$\beta$ & H$\gamma$        \\
  \hline
    10869.449 &  $5.932\pm 0.057$ & $3.980\pm 0.079$ &     --           \\               
    10875.449 &  $6.279\pm 0.087$ & $3.987\pm 0.093$ &     --           \\               
    10876.379 &  $6.365\pm 0.085$ & $4.082\pm 0.112$ &     --           \\               
    10905.395 &  $6.451\pm 0.057$ & $4.420\pm 0.080$ &     --           \\               
    10906.273 &  $6.352\pm 0.088$ & $4.329\pm 0.098$ &     --           \\               
    10924.432 &  $5.684\pm 0.079$ & $4.511\pm 0.085$ &     --           \\               
    10966.379 &  $5.008\pm 0.058$ & $3.476\pm 0.065$ &     --           \\               
    10982.395 &  $5.570\pm 0.101$ & $3.792\pm 0.096$ &     --           \\               
    10994.387 &  $5.678\pm 0.062$ & $4.072\pm 0.085$ &     --           \\               
    11014.516 &  $5.302\pm 0.091$ & $3.971\pm 0.094$ &     --           \\               
    11052.563 &  $5.519\pm 0.081$ & $3.724\pm 0.100$ &     --           \\               
    11074.574 &  $5.665\pm 0.064$ & $3.939\pm 0.086$ &     --           \\               
    11076.516 &  $5.628\pm 0.080$ & $4.022\pm 0.078$ &     --           \\               
    11085.570 &  $6.117\pm 0.066$ & $4.102\pm 0.087$ &     --           \\               
    11100.512 &  $6.344\pm 0.076$ & $4.283\pm 0.079$ &     --           \\               
    11141.406 &  $5.155\pm 0.087$ & $4.316\pm 0.087$ &     --           \\               
    11218.281 &  $5.525\pm 0.117$ & $4.022\pm 0.133$ &     --           \\               
    11278.395 &  $6.743\pm 0.074$ & $4.795\pm 0.103$ &     --           \\               
    11281.383 &  $6.927\pm 0.087$ & $4.687\pm 0.099$ &     --           \\               
    11290.402 &  $5.878\pm 0.187$ & $4.668\pm 0.201$ &     --           \\               
    11310.289 &  $5.216\pm 0.067$ & $4.594\pm 0.090$ &     --           \\               
    11319.313 &  $4.817\pm 0.063$ & $4.276\pm 0.086$ &     --           \\               
    11322.305 &  $5.134\pm 0.148$ & $3.885\pm 0.113$ &     --           \\               
    11349.324 &  $4.884\pm 0.084$ & $3.615\pm 0.078$ &     --           \\               
    11485.504 &  $5.096\pm 0.072$ & $4.438\pm 0.089$ &     --           \\               
    11497.434 &  $5.160\pm 0.086$ & $4.141\pm 0.113$ &     --           \\               
    11516.551 &  $5.103\pm 0.073$ & $4.020\pm 0.078$ &     --           \\               
    11557.418 &  $4.465\pm 0.077$ & $3.188\pm 0.098$ &     --           \\               
    11577.406 &  $4.456\pm 0.066$ & $3.151\pm 0.087$ &     --           \\               
    11587.293 &  $4.514\pm 0.070$ & $3.119\pm 0.079$ &     --           \\               
    11606.383 &  $4.821\pm 0.112$ & $3.192\pm 0.127$ &     --           \\               
    11608.285 &  $4.540\pm 0.089$ & $3.187\pm 0.094$ &     --           \\               
    11615.340 &  $5.173\pm 0.114$ & $3.081\pm 0.108$ &     --           \\               
    11636.270 &  $5.252\pm 0.076$ & $3.561\pm 0.084$ &     --           \\               
    11707.402 &  $5.783\pm 0.095$ & $3.687\pm 0.092$ &     --           \\               
    11720.336 &  $5.875\pm 0.069$ & $3.671\pm 0.087$ &     --           \\               
    11780.555 &  $5.757\pm 0.102$ & $4.052\pm 0.123$ &     --           \\               
    11782.555 &  $5.868\pm 0.080$ & $3.989\pm 0.101$ &     --           \\               
    11791.531 &  $5.385\pm 0.093$ & $4.017\pm 0.126$ &     --           \\               
    11810.531 &  $5.463\pm 0.075$ & $3.751\pm 0.097$ &     --           \\               
    11821.582 &  $5.010\pm 0.092$ & $4.032\pm 0.094$ &     --           \\               
    11823.496 &  $5.049\pm 0.072$ & $4.068\pm 0.097$ &     --           \\               
    11838.508 &  $5.234\pm 0.183$ & $3.741\pm 0.181$ &     --           \\               
    11840.539 &  $5.027\pm 0.103$ & $3.574\pm 0.111$ &     --           \\               
    11844.496 &  $4.679\pm 0.072$ & $3.577\pm 0.098$ &     --           \\               
    11847.473 &  $4.688\pm 0.089$ & $3.433\pm 0.116$ &     --           \\               
    11853.582 &  $4.628\pm 0.067$ & $3.456\pm 0.089$ &     --           \\               
    11869.496 &  $5.184\pm 0.073$ & $3.363\pm 0.082$ &     --           \\               
    11878.449 &  $5.012\pm 0.080$ & $3.450\pm 0.102$ &     --           \\               
    11901.316 &  $5.436\pm 0.106$ & $3.354\pm 0.123$ &     --           \\               
    11926.270 &  $4.391\pm 0.084$ & $3.374\pm 0.080$ &     --           \\               
    11959.227 &  $4.689\pm 0.129$ & $3.360\pm 0.151$ &     --           \\               
    11999.270 &  $4.517\pm 0.112$ & $3.077\pm 0.103$ &     --           \\               
    12030.422 &  $5.191\pm 0.100$ & $3.149\pm 0.105$ &     --           \\               
    12050.465 &  $4.689\pm 0.087$ & $3.219\pm 0.081$ &     --           \\               
    12104.316 &  $4.394\pm 0.069$ & $2.805\pm 0.079$ &     --           \\               
    12174.539 &  $4.327\pm 0.069$ & $3.007\pm 0.087$ &     --           \\               
    12175.531 &  $4.153\pm 0.084$ & $3.094\pm 0.087$ &     --           \\               
    12192.551 &  $4.046\pm 0.095$ & $3.099\pm 0.104$ &     --           \\               
    12201.598 &  $3.845\pm 0.105$ & $2.994\pm 0.133$ &     --           \\               
    12223.543 &  $4.004\pm 0.085$ & $ 2.934\pm 0.107$ &      --           \\                 
    12323.375 &  $3.899\pm 0.054$ & $ 2.424\pm 0.063$ &      --           \\             
    12343.473 &  $3.573\pm 0.107$ & $ 2.723\pm 0.118$ &      --           \\             
    
  \hline
\end{tabular}
\end{table}
\begin{table}
  \contcaption{F5170 spectral continuum, the H$\beta$ and H$\gamma$ line fluxes.}
  \begin{tabular}{@{}cccc@{}}
  \hline
JD-2,440,000& $F5170$   & H$\beta$ & H$\gamma$   \\
  \hline
    12381.262 &  $4.039\pm 0.125$ & $ 2.356\pm 0.130$ &      --           \\             
    12441.359 &  $5.067\pm 0.078$ & $ 3.406\pm 0.082$ &      --           \\             
    12613.543 &  $5.260\pm 0.085$ & $ 3.500\pm 0.099$ &      --           \\             
    12619.543 &  $5.285\pm 0.063$ & $ 3.457\pm 0.075$ &      --           \\             
    12675.410 &  $5.032\pm 0.106$ & $ 2.854\pm 0.126$ &      --           \\             
    12914.605 &  $5.412\pm 0.152$ & $ 3.171\pm 0.156$ &      --           \\             
    12915.531 &  $5.223\pm 0.067$ & $ 3.445\pm 0.092$ &      --           \\             
    13089.324 &  $4.197\pm 0.062$ & $ 2.682\pm 0.066$ &      --           \\             
    13097.301 &  $4.849\pm 0.205$ & $ 2.603\pm 0.111$ &      --           \\             
    13319.496 &  $3.639\pm 0.065$ & $ 2.741\pm 0.055$ &      --           \\             
    13356.555 &  $3.175\pm 0.060$ & $ 2.219\pm 0.068$ &      --           \\             
    13611.563 &  $6.791\pm 0.081$ & $ 4.125\pm 0.075$ & $ 1.342\pm 0.057$ \\
    13612.553 &  $6.812\pm 0.062$ & $ 4.135\pm 0.058$ & $ 1.368\pm 0.046$ \\
    13621.523 &  $6.877\pm 0.082$ & $ 4.264\pm 0.079$ & $ 1.417\pm 0.064$ \\
    13641.566 &  $6.279\pm 0.096$ & $ 4.197\pm 0.078$ & $ 1.347\pm 0.069$ \\
    13647.570 &  $6.158\pm 0.095$ & $ 4.157\pm 0.075$ & $ 1.260\pm 0.062$ \\
    13648.594 &  $6.310\pm 0.084$ & $ 4.156\pm 0.077$ & $ 1.332\pm 0.069$ \\
    13649.559 &  $6.276\pm 0.081$ & $ 4.082\pm 0.073$ & $ 1.277\pm 0.063$ \\
    13670.555 &  $6.284\pm 0.083$ & $ 4.020\pm 0.068$ & $ 1.317\pm 0.060$ \\
    13676.583 &  $6.309\pm 0.079$ & $ 3.914\pm 0.065$ & $ 1.235\pm 0.055$ \\
    13698.469 &  $6.366\pm 0.103$ & $ 4.038\pm 0.081$ & $ 1.255\pm 0.067$ \\
    13774.453 &  $5.677\pm 0.189$ & $ 4.203\pm 0.114$ & $ 1.290\pm 0.104$ \\
    13788.344 &  $5.997\pm 0.091$ & $ 4.044\pm 0.067$ & $ 1.269\pm 0.058$ \\
    13830.238 &  $6.715\pm 0.113$ & $ 3.949\pm 0.086$ & $ 1.206\pm 0.070$ \\
    13875.332 &  $7.178\pm 0.086$ & $ 4.282\pm 0.086$ & $ 1.264\pm 0.071$ \\
    13994.566 &  $7.272\pm 0.080$ & $ 4.269\pm 0.077$ & $ 1.326\pm 0.062$ \\
    14013.613 &  $7.496\pm 0.089$ & $ 4.407\pm 0.087$ & $ 1.317\pm 0.066$ \\
    14038.555 &  $7.741\pm 0.089$ & $ 4.177\pm 0.086$ & $ 1.227\pm 0.073$ \\
    14048.566 &  $7.806\pm 0.136$ & $ 4.069\pm 0.100$ & $ 1.085\pm 0.082$ \\
    14064.551 &  $7.001\pm 0.085$ & $ 4.244\pm 0.081$ & $ 1.116\pm 0.057$ \\
    14065.617 &  $6.963\pm 0.079$ & $ 4.139\pm 0.075$ & $ 1.195\pm 0.063$ \\
    14076.461 &  $6.903\pm 0.156$ & $ 4.044\pm 0.082$ & $ 1.194\pm 0.076$ \\
    14078.551 &  $6.923\pm 0.101$ & $ 3.989\pm 0.090$ & $ 1.015\pm 0.072$ \\
    14086.559 &  $7.182\pm 0.097$ & $ 3.924\pm 0.080$ & $ 1.062\pm 0.065$ \\
    14106.531 &  $7.897\pm 0.243$ & $ 4.107\pm 0.186$ &         --        \\
    14111.473 &  $7.469\pm 0.079$ & $ 4.316\pm 0.078$ & $ 1.241\pm 0.064$ \\
    14142.594 &  $7.459\pm 0.097$ & $ 4.418\pm 0.090$ & $ 1.120\pm 0.074$ \\
    14156.270 &  $7.366\pm 0.132$ & $ 4.429\pm 0.102$ & $ 1.220\pm 0.077$ \\
    14331.543 &  $8.416\pm 0.086$ & $ 5.376\pm 0.095$ & $ 1.647\pm 0.073$ \\
    14369.516 &  $7.978\pm 0.124$ & $ 4.843\pm 0.094$ & $ 1.556\pm 0.081$ \\
    14370.539 &  $7.390\pm 0.197$ & $ 4.736\pm 0.093$ & $ 1.335\pm 0.074$ \\
    14377.582 &  $7.570\pm 0.110$ & $ 4.714\pm 0.107$ & $ 1.410\pm 0.088$ \\
    14386.457 &  $7.490\pm 0.098$ & $ 4.184\pm 0.097$ & $ 1.294\pm 0.082$ \\
    14391.551 &  $6.816\pm 0.097$ & $ 4.384\pm 0.076$ & $ 1.327\pm 0.061$ \\
    14392.539 &  $7.115\pm 0.097$ & $ 4.410\pm 0.081$ & $ 1.331\pm 0.066$ \\
    14393.504 &  $7.178\pm 0.093$ & $ 4.517\pm 0.092$ & $ 1.396\pm 0.071$ \\
    14423.559 &  $6.261\pm 0.101$ & $ 3.715\pm 0.088$ & $ 1.149\pm 0.084$ \\
    14438.438 &  $5.934\pm 0.100$ & $ 3.612\pm 0.075$ & $ 1.043\pm 0.066$ \\
    14480.414 &  $7.435\pm 0.082$ & $ 4.110\pm 0.080$ & $ 1.166\pm 0.067$ \\
    14481.398 &  $7.342\pm 0.086$ & $ 4.108\pm 0.083$ & $ 1.049\pm 0.065$ \\
    14482.434 &  $7.498\pm 0.091$ & $ 4.016\pm 0.086$ & $ 1.179\pm 0.075$ \\
    14496.367 &  $7.517\pm 0.114$ & $ 4.244\pm 0.094$ & $ 1.209\pm 0.074$ \\
    14497.418 &  $7.561\pm 0.123$ & $ 3.955\pm 0.099$ & $ 1.155\pm 0.082$ \\
    14499.328 &  $7.615\pm 0.090$ & $ 4.125\pm 0.088$ & $ 1.168\pm 0.077$ \\
    14508.418 &  $7.071\pm 0.106$ & $ 4.272\pm 0.088$ & $ 1.258\pm 0.076$ \\
    14537.340 &  $6.605\pm 0.099$ & $ 3.742\pm 0.086$ & $ 1.233\pm 0.082$ \\
    14554.324 &  $5.749\pm 0.125$ & $ 3.736\pm 0.099$ & $ 0.942\pm 0.083$ \\
    14574.309 &  $6.132\pm 0.269$ & $ 3.303\pm 0.226$ &         --        \\        
    14586.309 &  $6.284\pm 0.086$ & $ 3.550\pm 0.074$ & $ 1.064\pm 0.069$ \\    
    14587.281 &  $6.184\pm 0.092$ & $ 3.576\pm 0.080$ & $ 1.003\pm 0.067$ \\
    14588.266 &  $6.264\pm 0.098$ & $ 3.577\pm 0.082$ & $ 1.027\pm 0.068$ \\
    14590.309 &  $6.278\pm 0.098$ & $ 3.543\pm 0.076$ & $ 0.968\pm 0.060$ \\    
  \hline
\end{tabular}
\end{table}

\begin{table}
  \contcaption{F5170 spectral continuum, the H$\beta$ and H$\gamma$ line fluxes.}
  \begin{tabular}{@{}cccc}
  \hline
JD-2,440,000& $F5170$   & H$\beta$ & H$\gamma$        \\
  \hline
    14591.273 &  $6.440\pm 0.108$ & $ 3.556\pm 0.086$ & $ 1.020\pm 0.077$ \\
    14617.344 &  $7.076\pm 0.086$ & $ 3.632\pm 0.081$ & $ 0.955\pm 0.062$ \\
    14620.301 &  $7.053\pm 0.091$ & $ 3.923\pm 0.090$ & $ 1.053\pm 0.071$ \\
    14622.426 &  $7.141\pm 0.087$ & $ 3.682\pm 0.084$ & $ 1.048\pm 0.068$ \\
    14687.547 &  $6.498\pm 0.095$ & $ 4.099\pm 0.073$ & $ 1.091\pm 0.057$ \\
    14779.629 &  $7.584\pm 0.127$ & $ 4.711\pm 0.106$ & $ 1.328\pm 0.070$ \\
    14804.566 &  $8.214\pm 0.093$ & $ 4.650\pm 0.086$ & $ 1.305\pm 0.070$ \\
\hline
\end{tabular}
{\\\footnotesize 
Units are $10^{-13}\,\ergs$ and $10^{-15}\,\ergsA$ for the
lines and continuum, respectively.
}
\end{table}

\begin{table}
  \caption{F7030 spectral continuum and H$\alpha$ fluxes.}
  \label{haflx}
  \begin{tabular}{@{}ccc}
  \hline
JD-2,440,000 & $F7030$   & H$\alpha$      \\
\hline
10869.359 & $ 6.858\pm 0.194$ & $31.959\pm 0.796$ \\
10876.453 & $ 7.063\pm 0.177$ & $32.400\pm 0.833$ \\
10905.332 & $ 7.403\pm 0.176$ & $33.322\pm 0.722$ \\
11053.559 & $ 6.227\pm 0.177$ & $30.750\pm 0.726$ \\
11075.582 & $ 6.408\pm 0.171$ & $30.532\pm 0.713$ \\
11220.219 & $ 6.271\pm 0.231$ & $29.815\pm 1.056$ \\
11279.324 & $ 7.851\pm 0.210$ & $37.009\pm 0.962$ \\
11321.301 & $ 5.697\pm 0.265$ & $33.947\pm 1.258$ \\
11517.488 & $ 5.327\pm 0.268$ & $31.821\pm 1.134$ \\
11587.355 & $ 5.174\pm 0.195$ & $27.152\pm 0.741$ \\
11638.250 & $ 6.202\pm 0.163$ & $29.285\pm 0.743$ \\
11707.344 & $ 6.400\pm 0.165$ & $30.001\pm 0.712$ \\
11721.324 & $ 6.689\pm 0.188$ & $32.688\pm 0.838$ \\
11823.543 & $ 6.115\pm 0.221$ & $32.067\pm 0.951$ \\
11847.531 & $ 5.611\pm 0.204$ & $30.121\pm 0.891$ \\
11878.512 & $ 6.134\pm 0.216$ & $30.642\pm 0.918$ \\
11902.461 & $ 6.325\pm 0.219$ & $30.126\pm 0.899$ \\
12000.254 & $ 4.590\pm 0.381$ & $28.123\pm 1.239$ \\
12031.402 & $ 6.159\pm 0.277$ & $27.928\pm 1.015$ \\
12105.324 & $ 5.290\pm 0.183$ & $28.077\pm 0.731$ \\
12176.520 & $ 4.917\pm 0.209$ & $27.194\pm 0.780$ \\
12193.508 & $ 4.927\pm 0.263$ & $28.147\pm 1.009$ \\
12225.438 & $ 3.968\pm 0.598$ & $28.053\pm 1.987$ \\
12323.457 & $ 4.401\pm 0.166$ & $22.116\pm 0.518$ \\
12343.551 & $ 4.813\pm 0.394$ & $25.054\pm 1.615$ \\
12613.594 & $ 5.715\pm 0.277$ & $26.821\pm 1.083$ \\
12675.477 & $ 5.779\pm 0.193$ & $26.451\pm 0.827$ \\
13090.277 & $ 4.505\pm 0.297$ & $24.783\pm 0.909$ \\
13356.465 & $ 3.857\pm 0.159$ & $20.999\pm 0.656$ \\
13610.559 & $ 8.018\pm 0.215$ & $31.853\pm 0.796$ \\
13621.555 & $ 8.487\pm 0.240$ & $32.350\pm 0.891$ \\
13642.594 & $ 7.377\pm 0.235$ & $31.616\pm 1.047$ \\
13648.551 & $ 7.688\pm 0.277$ & $30.374\pm 1.081$ \\
13648.563 & $ 7.474\pm 0.218$ & $30.213\pm 0.859$ \\
13789.316 & $ 7.176\pm 0.227$ & $32.545\pm 0.986$ \\
13831.234 & $ 7.418\pm 0.186$ & $30.618\pm 0.825$ \\
13994.586 & $ 8.940\pm 0.228$ & $34.663\pm 0.875$ \\
14066.457 & $ 8.472\pm 0.257$ & $34.872\pm 1.008$ \\
14370.563 & $ 8.707\pm 0.264$ & $37.536\pm 1.117$ \\
14392.563 & $ 8.377\pm 0.206$ & $36.161\pm 0.988$ \\
14393.527 & $ 8.125\pm 0.256$ & $35.540\pm 1.017$ \\
14481.387 & $ 8.255\pm 0.243$ & $34.970\pm 1.044$ \\
14508.445 & $ 8.117\pm 0.377$ & $36.963\pm 1.866$ \\
14617.324 & $ 7.455\pm 0.205$ & $31.689\pm 0.795$ \\
14620.316 & $ 8.052\pm 0.228$ & $33.558\pm 0.900$ \\
14622.367 & $ 7.785\pm 0.204$ & $32.227\pm 0.890$ \\
14804.543 & $10.004\pm 0.242$ & $38.158\pm 0.965$ \\
\hline
\end{tabular}
{\\\footnotesize 
Units are $10^{-13}\,\ergs$ and $10^{-15}\,\ergsA$ for the
\ha\ line and continuum, respectively. \\
}
\end{table}

\begin{table}
  \caption{Combined F5170 continuum fluxes from spectral and photometric observations.}
  \label{cntflx}
  \begin{tabular}{@{}lclc}
  \hline
JD-2,440,000 & $F5170$ &Julian Date & $F5170$             \\
\hline
   8630.5781   &  $4.072\pm 0.107$ &    9783.2969   &  $7.666\pm 0.098$   \\
   8716.4258   &  $5.003\pm 0.143$ &    9814.2656   &  $7.775\pm 0.124$   \\
   8717.3633   &  $5.083\pm 0.129$ &    9838.3027   &  $7.282\pm 0.356$   \\
   8927.5117   &  $4.125\pm 0.054$ &    9839.3398   &  $6.827\pm 0.240$   \\
   8983.4009   &  $4.277\pm 0.050$ &    9867.3389   &  $7.121\pm 0.303$   \\
   9001.2852   &  $4.483\pm 0.151$ &    9871.3545   &  $6.683\pm 0.252$   \\
   9031.2314   &  $3.991\pm 0.150$ &    9872.3594   &  $6.775\pm 0.080$   \\
   9057.2441   &  $4.225\pm 0.146$ &    9980.5746   &  $6.749\pm 0.058$   \\
   9059.3291   &  $4.483\pm 0.171$ &   10008.5636   &  $7.303\pm 0.052$   \\
   9062.3330   &  $4.331\pm 0.161$ &   10009.5860   &  $7.380\pm 0.072$   \\
   9070.2891   &  $4.274\pm 0.068$ &   10010.5900   &  $7.445\pm 0.083$   \\
   9074.3203   &  $4.378\pm 0.059$ &   10013.5390   &  $7.236\pm 0.251$   \\
   9088.2656   &  $5.198\pm 0.076$ &   10015.5170   &  $7.458\pm 0.255$   \\
   9089.2578   &  $5.019\pm 0.079$ &   10024.5360   &  $7.791\pm 0.289$   \\
   9100.2910   &  $5.021\pm 0.186$ &   10036.5900   &  $7.798\pm 0.119$   \\
   9101.2930   &  $5.116\pm 0.174$ &   10047.5526   &  $8.101\pm 0.057$   \\
   9141.3047   &  $6.684\pm 0.119$ &   10064.5120   &  $7.593\pm 0.072$   \\
   9156.3203   &  $7.048\pm 0.126$ &   10069.4550   &  $6.782\pm 0.231$   \\
   9250.5977   &  $6.075\pm 0.170$ &   10092.2260   &  $7.053\pm 0.241$   \\
   9252.5757   &  $6.082\pm 0.093$ &   10094.2990   &  $7.399\pm 0.253$   \\
   9255.4814   &  $5.841\pm 0.210$ &   10096.3280   &  $6.939\pm 0.234$   \\
   9272.6016   &  $5.799\pm 0.068$ &   10102.3720   &  $7.282\pm 0.255$   \\
   9273.4551   &  $5.769\pm 0.211$ &   10133.2640   &  $6.606\pm 0.229$   \\
   9274.5796   &  $5.568\pm 0.047$ &   10135.3040   &  $7.540\pm 0.276$   \\
   9275.4990   &  $5.164\pm 0.184$ &   10139.4580   &  $7.376\pm 0.252$   \\
   9311.5742   &  $5.665\pm 0.077$ &   10156.3134   &  $7.602\pm 0.067$   \\
   9313.4385   &  $5.567\pm 0.061$ &   10159.3190   &  $7.949\pm 0.275$   \\
   9329.4219   &  $5.841\pm 0.214$ &   10161.4450   &  $7.790\pm 0.082$   \\
   9331.4531   &  $6.418\pm 0.068$ &   10201.5550   &  $6.994\pm 0.165$   \\
   9332.4521   &  $6.027\pm 0.212$ &   10202.4060   &  $7.008\pm 0.092$   \\
   9341.3809   &  $6.313\pm 0.219$ &   10212.2930   &  $5.893\pm 0.229$   \\
   9357.3135   &  $6.540\pm 0.233$ &   10213.4450   &  $6.383\pm 0.078$   \\
   9359.3467   &  $6.090\pm 0.211$ &   10218.3130   &  $5.790\pm 0.225$   \\
   9362.3525   &  $6.017\pm 0.208$ &   10222.3060   &  $6.101\pm 0.223$   \\
   9364.3281   &  $6.334\pm 0.067$ &   10225.3320   &  $5.847\pm 0.077$   \\
   9365.4180   &  $6.448\pm 0.060$ &   10246.3710   &  $5.474\pm 0.060$   \\
   9395.3750   &  $6.140\pm 0.068$ &   10258.3980   &  $5.091\pm 0.110$   \\
   9399.3984   &  $6.155\pm 0.088$ &   10304.4220   &  $6.479\pm 0.081$   \\
   9450.3438   &  $6.618\pm 0.062$ &   10361.5310   &  $6.989\pm 0.068$   \\
   9452.2461   &  $6.651\pm 0.100$ &   10363.4840   &  $7.053\pm 0.070$   \\
   9454.2734   &  $6.731\pm 0.077$ &   10364.5410   &  $7.064\pm 0.274$   \\
   9488.3398   &  $7.461\pm 0.072$ &   10372.4804   &  $7.427\pm 0.040$   \\
   9520.3281   &  $6.829\pm 0.076$ &   10392.5060   &  $7.411\pm 0.271$   \\
   9536.3867   &  $6.575\pm 0.064$ &   10395.4687   &  $7.013\pm 0.049$   \\
   9548.5000   &  $6.048\pm 0.099$ &   10396.4610   &  $7.103\pm 0.063$   \\
   9554.5234   &  $5.900\pm 0.111$ &   10397.4610   &  $7.274\pm 0.063$   \\
   9555.5117   &  $5.813\pm 0.124$ &   10399.4490   &  $7.179\pm 0.078$   \\
   9566.4570   &  $5.304\pm 0.068$ &   10401.4600   &  $7.018\pm 0.246$   \\
   9578.3984   &  $5.070\pm 0.058$ &   10403.4010   &  $7.599\pm 0.267$   \\
   9599.3672   &  $4.653\pm 0.134$ &   10404.4920   &  $7.635\pm 0.258$   \\
   9622.2734   &  $5.361\pm 0.094$ &   10406.5080   &  $7.446\pm 0.240$   \\
   9639.5547   &  $5.878\pm 0.096$ &   10408.5900   &  $7.202\pm 0.229$   \\
   9653.5635   &  $6.298\pm 0.074$ &   10430.3770   &  $6.573\pm 0.224$   \\
   9665.5331   &  $5.784\pm 0.069$ &   10434.3200   &  $6.386\pm 0.055$   \\
   9685.3984   &  $6.646\pm 0.147$ &   10435.3240   &  $6.586\pm 0.072$   \\
   9691.5078   &  $6.915\pm 0.086$ &   10436.4100   &  $6.371\pm 0.052$   \\
   9713.4539   &  $7.374\pm 0.052$ &   10461.5560   &  $5.841\pm 0.195$   \\
   9716.5488   &  $7.064\pm 0.252$ &   10482.5040   &  $6.621\pm 0.088$   \\
   9723.4297   &  $7.226\pm 0.081$ &   10483.4790   &  $6.683\pm 0.220$   \\
   9744.3984   &  $8.089\pm 0.078$ &   10484.4480   &  $6.628\pm 0.209$   \\
   9753.3086   &  $8.467\pm 0.074$ &   10487.3660   &  $6.410\pm 0.213$   \\
   9754.3789   &  $8.249\pm 0.183$ &   10491.4380   &  $6.573\pm 0.222$   \\

   \hline
\end{tabular}
\end{table}

\begin{table}
  \contcaption{Combined F5170 continuum fluxes from spectral and photometric observations.}
  \begin{tabular}{@{}lclc}
  \hline
JD-2,440,000 & $F5170$  &Julian Date & $F5170$               \\
\hline
   9771.3086   &  $7.988\pm 0.071$ &   10495.3670   &  $6.383\pm 0.057$   \\
  10509.3810  &  $6.054\pm 0.062$ &   11085.5700   &  $6.117\pm 0.066$   \\
  10510.4262  &  $5.990\pm 0.044$ &   11088.4720   &  $5.330\pm 0.184$   \\
  10511.3540  &  $5.608\pm 0.196$ &   11100.5120   &  $6.344\pm 0.076$   \\
  10518.4320  &  $5.408\pm 0.176$ &   11105.5480   &  $6.584\pm 0.214$   \\
  10519.4260  &  $5.638\pm 0.190$ &   11110.5780   &  $5.790\pm 0.195$   \\
  10521.3830  &  $5.709\pm 0.192$ &   11111.5760   &  $5.924\pm 0.207$   \\
  10522.3851  &  $5.517\pm 0.067$ &   11141.4334   &  $5.127\pm 0.081$   \\
  10541.3549  &  $5.524\pm 0.046$ &   11163.4850   &  $5.729\pm 0.185$   \\
  10543.2960  &  $5.059\pm 0.188$ &   11164.3130   &  $6.207\pm 0.220$   \\
  10566.3030  &  $6.464\pm 0.230$ &   11176.5790   &  $5.478\pm 0.192$   \\
  10569.3380  &  $5.517\pm 0.196$ &   11192.3790   &  $5.810\pm 0.201$   \\
  10574.3634  &  $5.080\pm 0.038$ &   11197.2630   &  $5.934\pm 0.211$   \\
  10575.3160  &  $4.840\pm 0.054$ &   11199.3480   &  $5.021\pm 0.174$   \\
  10576.3520  &  $4.868\pm 0.058$ &   11218.2810   &  $5.525\pm 0.117$   \\
  10580.3820  &  $5.252\pm 0.210$ &   11261.4920   &  $7.018\pm 0.250$   \\
  10597.3852  &  $5.460\pm 0.083$ &   11274.3120   &  $6.951\pm 0.229$   \\
  10601.3550  &  $5.540\pm 0.064$ &   11278.3950   &  $6.743\pm 0.074$   \\
  10611.3440  &  $6.296\pm 0.099$ &   11279.2870   &  $6.650\pm 0.237$   \\
  10628.3590  &  $6.778\pm 0.081$ &   11281.3662   &  $6.943\pm 0.081$   \\
  10642.3320  &  $6.824\pm 0.068$ &   11290.4020   &  $5.878\pm 0.187$   \\
  10654.3320  &  $6.808\pm 0.153$ &   11306.3890   &  $5.709\pm 0.203$   \\
  10655.3280  &  $6.764\pm 0.081$ &   11310.2890   &  $5.216\pm 0.067$   \\
  10687.5700  &  $6.033\pm 0.116$ &   11319.3130   &  $4.817\pm 0.063$   \\
  10697.4208  &  $5.833\pm 0.048$ &   11322.3050   &  $5.134\pm 0.148$   \\
  10699.5270  &  $5.661\pm 0.057$ &   11346.3850   &  $5.164\pm 0.203$   \\
  10705.5590  &  $5.587\pm 0.188$ &   11349.3240   &  $4.884\pm 0.084$   \\
  10714.5080  &  $5.504\pm 0.063$ &   11400.5350   &  $4.851\pm 0.247$   \\
  10715.5200  &  $5.526\pm 0.061$ &   11407.5290   &  $5.021\pm 0.167$   \\
  10728.5390  &  $5.333\pm 0.052$ &   11409.5370   &  $4.841\pm 0.191$   \\
  10729.5080  &  $5.399\pm 0.056$ &   11454.5560   &  $5.658\pm 0.207$   \\
  10747.5430  &  $5.962\pm 0.076$ &   11467.4890   &  $5.310\pm 0.181$   \\
  10748.4650  &  $5.950\pm 0.057$ &   11485.5040   &  $5.096\pm 0.072$   \\
  10755.5340  &  $6.196\pm 0.211$ &   11488.6040   &  $5.136\pm 0.193$   \\
  10758.5249  &  $5.924\pm 0.052$ &   11493.3450   &  $5.438\pm 0.202$   \\
  10759.4380  &  $5.777\pm 0.076$ &   11497.4340   &  $5.160\pm 0.086$   \\
  10760.5280  &  $5.924\pm 0.195$ &   11516.5510   &  $5.103\pm 0.073$   \\
  10761.5130  &  $6.185\pm 0.206$ &   11522.5520   &  $4.610\pm 0.162$   \\
  10762.5670  &  $6.196\pm 0.209$ &   11524.5720   &  $5.184\pm 0.231$   \\
  10777.4300  &  $5.776\pm 0.067$ &   11525.4560   &  $5.126\pm 0.190$   \\
  10801.2660  &  $5.955\pm 0.203$ &   11557.3624   &  $4.581\pm 0.056$   \\
  10817.3570  &  $5.389\pm 0.177$ &   11577.4060   &  $4.456\pm 0.066$   \\
  10863.3710  &  $6.562\pm 0.221$ &   11581.3890   &  $4.547\pm 0.153$   \\
  10866.3790  &  $6.377\pm 0.234$ &   11586.4480   &  $4.601\pm 0.157$   \\
  10867.4230  &  $6.291\pm 0.210$ &   11587.3295   &  $4.542\pm 0.053$   \\
  10868.4180  &  $6.453\pm 0.212$ &   11588.5210   &  $4.907\pm 0.163$   \\
  10869.4520  &  $5.996\pm 0.047$ &   11598.2560   &  $4.721\pm 0.180$   \\
  10873.4040  &  $6.410\pm 0.208$ &   11603.2890   &  $4.879\pm 0.192$   \\
  10874.2920  &  $6.324\pm 0.228$ &   11605.4510   &  $5.145\pm 0.191$   \\
  10875.4490  &  $6.279\pm 0.209$ &   11606.3830   &  $4.821\pm 0.112$   \\
  10876.3961  &  $6.207\pm 0.063$ &   11608.3236   &  $4.493\pm 0.056$   \\
  10905.3950  &  $6.451\pm 0.057$ &   11612.3560   &  $5.097\pm 0.181$   \\
  10906.2730  &  $6.352\pm 0.088$ &   11615.3400   &  $5.173\pm 0.114$   \\
  10924.4482  &  $5.708\pm 0.074$ &   11628.3200   &  $5.729\pm 0.209$   \\
  10957.3660  &  $5.136\pm 0.200$ &   11636.2700   &  $5.252\pm 0.076$   \\
  10966.3790  &  $5.008\pm 0.058$ &   11661.2870   &  $5.310\pm 0.194$   \\
  10982.3950  &  $5.570\pm 0.101$ &   11707.4020   &  $5.783\pm 0.095$   \\
  10994.3922  &  $5.671\pm 0.053$ &   11720.3360   &  $5.875\pm 0.069$   \\
  11014.5160  &  $5.302\pm 0.091$ &   11780.5550   &  $5.757\pm 0.102$   \\
  11044.3860  &  $5.408\pm 0.220$ &   11782.5550   &  $5.868\pm 0.080$   \\
  11050.5250  &  $5.498\pm 0.181$ &   11788.5360   &  $5.638\pm 0.243$   \\
  11052.5630  &  $5.519\pm 0.081$ &   11791.5310   &  $5.385\pm 0.093$   \\
\hline
\end{tabular}
\end{table}

\begin{table}
  \contcaption{Combined F5170 continuum fluxes from spectral and photometric observations.}
  \begin{tabular}{@{}lclc}
  \hline
JD-2,440,000 & $F5170$    &Julian Date & $F5170$                   \\
\hline
  11074.5740  &  $5.665\pm 0.064$ &   11810.5310   &  $5.463\pm 0.075$   \\
  11076.4995  &  $5.650\pm 0.077$ &   11817.5320   &  $5.986\pm 0.243$   \\
  11818.5550   &  $5.408\pm 0.185$ & 12349.3890   &  $3.483\pm 0.133$  \\ 
  11821.5820   &  $5.010\pm 0.092$ & 12366.2620   &  $3.890\pm 0.024$  \\
  11823.4960   &  $5.049\pm 0.072$ & 12367.2780   &  $3.984\pm 0.026$  \\
  11838.5080   &  $5.234\pm 0.183$ & 12368.3450   &  $4.077\pm 0.057$  \\
  11840.5390   &  $5.027\pm 0.103$ & 12369.2769   &  $4.188\pm 0.022$  \\
  11842.4730   &  $5.388\pm 0.206$ & 12381.2620   &  $4.039\pm 0.125$  \\
  11843.5660   &  $5.184\pm 0.439$ & 12385.3530   &  $4.388\pm 0.033$  \\
  11844.4960   &  $4.679\pm 0.072$ & 12386.2840   &  $4.481\pm 0.038$  \\
  11847.5150   &  $4.600\pm 0.062$ & 12387.2960   &  $4.401\pm 0.034$  \\
  11853.5681   &  $4.767\pm 0.049$ & 12388.3290   &  $4.504\pm 0.050$  \\
  11866.4100   &  $5.478\pm 0.233$ & 12399.3050   &  $4.898\pm 0.169$  \\
  11867.3210   &  $5.155\pm 0.197$ & 12403.3180   &  $5.097\pm 0.190$  \\
  11868.3710   &  $5.349\pm 0.190$ & 12404.2690   &  $5.073\pm 0.029$  \\
  11869.4960   &  $5.184\pm 0.073$ & 12405.3055   &  $5.015\pm 0.030$  \\
  11878.4490   &  $5.012\pm 0.080$ & 12406.3010   &  $5.193\pm 0.227$  \\
  11879.2630   &  $5.222\pm 0.212$ & 12407.2730   &  $4.971\pm 0.034$  \\
  11882.4800   &  $4.656\pm 0.170$ & 12408.2942   &  $5.104\pm 0.025$  \\
  11901.3160   &  $5.436\pm 0.106$ & 12409.3480   &  $5.078\pm 0.216$  \\
  11902.2690   &  $5.688\pm 0.205$ & 12410.2870   &  $5.132\pm 0.028$  \\
  11912.5120   &  $5.116\pm 0.190$ & 12411.2860   &  $5.119\pm 0.029$  \\
  11926.2770   &  $4.424\pm 0.056$ & 12417.2820   &  $5.289\pm 0.052$  \\
  11932.4920   &  $4.795\pm 0.184$ & 12419.3020   &  $5.272\pm 0.049$  \\
  11959.2270   &  $4.689\pm 0.129$ & 12421.2880   &  $5.175\pm 0.058$  \\
  11999.2700   &  $4.517\pm 0.112$ & 12440.2860   &  $5.165\pm 0.050$  \\
  12030.4220   &  $5.191\pm 0.100$ & 12441.3590   &  $5.067\pm 0.078$  \\
  12050.4650   &  $4.689\pm 0.087$ & 12442.2850   &  $5.123\pm 0.044$  \\
  12104.3160   &  $4.394\pm 0.069$ & 12459.2830   &  $5.198\pm 0.064$  \\
  12139.5420   &  $4.138\pm 0.196$ & 12476.2920   &  $5.265\pm 0.063$  \\
  12144.5390   &  $4.620\pm 0.200$ & 12484.3370   &  $4.793\pm 0.064$  \\
  12147.5200   &  $4.060\pm 0.155$ & 12498.4950   &  $4.379\pm 0.024$  \\
  12166.5140   &  $4.629\pm 0.167$ & 12530.5830   &  $4.424\pm 0.031$  \\
  12174.5390   &  $4.327\pm 0.069$ & 12536.5580   &  $4.640\pm 0.025$  \\
  12175.5310   &  $4.153\pm 0.084$ & 12539.5520   &  $4.732\pm 0.040$  \\
  12192.5510   &  $4.046\pm 0.095$ & 12541.5710   &  $4.653\pm 0.041$  \\
  12199.5620   &  $4.094\pm 0.151$ & 12557.5940   &  $4.579\pm 0.027$  \\
  12201.5449   &  $3.888\pm 0.054$ & 12566.4570   &  $4.422\pm 0.036$  \\
  12210.5970   &  $4.129\pm 0.149$ & 12569.5400   &  $4.648\pm 0.043$  \\
  12223.5055   &  $4.090\pm 0.060$ & 12593.6361   &  $4.954\pm 0.035$  \\
  12225.3400   &  $4.340\pm 0.161$ & 12595.3070   &  $4.844\pm 0.035$  \\
  12231.5060   &  $4.060\pm 0.151$ & 12596.5600   &  $4.890\pm 0.043$  \\
  12263.4680   &  $4.094\pm 0.024$ & 12597.5780   &  $4.956\pm 0.044$  \\
  12265.5230   &  $4.182\pm 0.140$ & 12605.5350   &  $5.416\pm 0.051$  \\
  12280.5470   &  $3.783\pm 0.020$ & 12608.5690   &  $5.428\pm 0.034$  \\
  12281.4510   &  $3.725\pm 0.018$ & 12609.5610   &  $5.467\pm 0.030$  \\
  12283.4430   &  $3.801\pm 0.020$ & 12610.4815   &  $5.324\pm 0.028$  \\
  12298.4070   &  $3.623\pm 0.022$ & 12612.5340   &  $5.550\pm 0.030$  \\
  12301.3840   &  $3.625\pm 0.028$ & 12613.4293   &  $5.497\pm 0.030$  \\
  12307.2820   &  $3.427\pm 0.129$ & 12614.5590   &  $5.548\pm 0.033$  \\
  12308.3500   &  $3.628\pm 0.019$ & 12618.3620   &  $5.532\pm 0.026$  \\
  12309.4384   &  $3.599\pm 0.019$ & 12619.5547   &  $5.452\pm 0.026$  \\
  12310.4000   &  $3.307\pm 0.128$ & 12620.4070   &  $5.375\pm 0.035$  \\
  12313.4260   &  $3.307\pm 0.132$ & 12621.4710   &  $5.477\pm 0.034$  \\
  12314.3730   &  $3.823\pm 0.030$ & 12625.4390   &  $5.301\pm 0.036$  \\
  12316.4110   &  $3.906\pm 0.020$ & 12634.2880   &  $4.730\pm 0.164$  \\
  12321.4275   &  $4.018\pm 0.023$ & 12635.4457   &  $5.088\pm 0.155$  \\
  12322.4080   &  $4.144\pm 0.023$ & 12636.3986   &  $5.035\pm 0.042$  \\
  12323.3560   &  $4.108\pm 0.021$ & 12665.3620   &  $4.728\pm 0.026$  \\
  12324.3990   &  $4.152\pm 0.126$ & 12672.3350   &  $4.974\pm 0.038$  \\
  12336.2490   &  $3.507\pm 0.110$ & 12674.4100   &  $5.041\pm 0.031$  \\
  12342.4700   &  $3.402\pm 0.144$ & 12675.3846   &  $5.002\pm 0.028$  \\
\hline
\end{tabular}
\end{table}

\begin{table}
  \contcaption{Combined F5170 continuum fluxes from spectral and photometric observations.}
  \begin{tabular}{@{}lclc}
  \hline
JD-2,440,000 & $F5170$   &Julian Date & $F5170$            \\
\hline
  12343.4263   &  $3.559\pm 0.046$ & 12683.3540   &  $4.635\pm 0.037$  \\
  12346.3890   &  $3.499\pm 0.132$ & 12684.3840   &  $4.703\pm 0.034$  \\
  12348.2790   &  $3.315\pm 0.141$ & 12685.3890   &  $4.632\pm 0.051$  \\
  12689.3540   &  $4.830\pm 0.041$ & 12998.4690   &  $4.207\pm 0.024$  \\
  12694.3330   &  $4.772\pm 0.034$ & 13003.4349   &  $4.187\pm 0.022$  \\
  12696.3900   &  $5.087\pm 0.195$ & 13006.4560   &  $4.084\pm 0.056$  \\
  12697.3620   &  $4.836\pm 0.025$ & 13007.4330   &  $4.046\pm 0.034$  \\
  12698.3500   &  $4.913\pm 0.027$ & 13015.4930   &  $3.971\pm 0.057$  \\
  12700.3786   &  $4.836\pm 0.027$ & 13022.4180   &  $4.391\pm 0.024$  \\
  12701.3560   &  $4.781\pm 0.027$ & 13023.3250   &  $4.367\pm 0.143$  \\
  12703.2680   &  $4.711\pm 0.175$ & 13058.3160   &  $4.776\pm 0.026$  \\
  12710.2760   &  $4.595\pm 0.028$ & 13071.4320   &  $4.767\pm 0.072$  \\
  12716.2540   &  $4.167\pm 0.033$ & 13073.3260   &  $4.521\pm 0.033$  \\
  12722.3690   &  $4.069\pm 0.145$ & 13077.3130   &  $4.367\pm 0.136$  \\
  12723.4030   &  $3.889\pm 0.142$ & 13083.3410   &  $4.438\pm 0.030$  \\
  12724.3289   &  $4.009\pm 0.024$ & 13084.3190   &  $4.500\pm 0.029$  \\
  12726.2940   &  $3.771\pm 0.138$ & 13084.3190   &  $4.500\pm 0.029$  \\
  12727.3100   &  $4.033\pm 0.028$ & 13085.3510   &  $4.558\pm 0.028$  \\
  12728.3020   &  $4.123\pm 0.024$ & 13087.2780   &  $4.425\pm 0.028$  \\
  12729.3010   &  $4.177\pm 0.024$ & 13089.3240   &  $4.197\pm 0.062$  \\
  12730.2840   &  $4.232\pm 0.029$ & 13097.2904   &  $4.606\pm 0.041$  \\
  12739.2820   &  $4.235\pm 0.028$ & 13098.2500   &  $4.636\pm 0.038$  \\
  12740.3020   &  $4.185\pm 0.031$ & 13105.3080   &  $4.556\pm 0.152$  \\
  12742.3080   &  $4.205\pm 0.035$ & 13111.3030   &  $4.978\pm 0.027$  \\
  12744.3070   &  $4.221\pm 0.030$ & 13112.3115   &  $4.799\pm 0.025$  \\
  12745.2760   &  $4.272\pm 0.037$ & 13113.3120   &  $4.864\pm 0.028$  \\
  12751.2799   &  $4.242\pm 0.024$ & 13114.2551   &  $5.012\pm 0.030$  \\
  12752.2510   &  $4.309\pm 0.024$ & 13115.2890   &  $5.108\pm 0.029$  \\
  12754.2770   &  $4.383\pm 0.024$ & 13117.3010   &  $4.814\pm 0.164$  \\
  12756.2860   &  $4.324\pm 0.024$ & 13130.2730   &  $5.063\pm 0.044$  \\
  12757.2770   &  $4.296\pm 0.024$ & 13133.2930   &  $5.075\pm 0.026$  \\
  12759.2830   &  $4.647\pm 0.165$ & 13135.2730   &  $5.004\pm 0.029$  \\
  12766.2680   &  $4.454\pm 0.025$ & 13148.2890   &  $4.770\pm 0.030$  \\
  12767.2590   &  $4.544\pm 0.029$ & 13149.2750   &  $4.727\pm 0.045$  \\
  12770.2590   &  $4.320\pm 0.038$ & 13153.2910   &  $4.802\pm 0.032$  \\
  12771.2560   &  $4.272\pm 0.051$ & 13154.3280   &  $4.884\pm 0.052$  \\
  12774.2750   &  $4.356\pm 0.035$ & 13277.5570   &  $5.045\pm 0.079$  \\
  12775.2710   &  $4.353\pm 0.043$ & 13291.5610   &  $4.765\pm 0.029$  \\
  12778.2980   &  $4.492\pm 0.185$ & 13292.5770   &  $4.817\pm 0.029$  \\
  12790.3080   &  $4.933\pm 0.028$ & 13296.6160   &  $4.355\pm 0.027$  \\
  12791.2820   &  $4.859\pm 0.040$ & 13300.6100   &  $4.245\pm 0.029$  \\
  12841.5450   &  $4.824\pm 0.104$ & 13302.6060   &  $4.161\pm 0.029$  \\
  12866.5290   &  $5.115\pm 0.034$ & 13305.6230   &  $4.073\pm 0.027$  \\
  12883.5410   &  $5.011\pm 0.171$ & 13307.6120   &  $4.107\pm 0.053$  \\
  12889.5360   &  $5.408\pm 0.169$ & 13308.6250   &  $4.040\pm 0.046$  \\
  12890.5250   &  $5.376\pm 0.032$ & 13309.6280   &  $3.986\pm 0.039$  \\
  12903.5440   &  $5.300\pm 0.186$ & 13313.6020   &  $4.011\pm 0.027$  \\
  12906.5980   &  $5.109\pm 0.031$ & 13314.5960   &  $4.109\pm 0.027$  \\
  12907.5800   &  $4.926\pm 0.159$ & 13315.6030   &  $4.046\pm 0.024$  \\
  12912.5813   &  $5.089\pm 0.030$ & 13317.5910   &  $3.993\pm 0.027$  \\
  12913.5490   &  $5.132\pm 0.034$ & 13318.6095   &  $3.912\pm 0.028$  \\
  12914.5768   &  $5.108\pm 0.028$ & 13319.4960   &  $3.639\pm 0.065$  \\
  12915.5268   &  $5.260\pm 0.049$ & 13320.4660   &  $3.581\pm 0.116$  \\
  12945.6380   &  $4.830\pm 0.047$ & 13323.5880   &  $3.991\pm 0.133$  \\
  12947.5950   &  $4.664\pm 0.032$ & 13331.5011   &  $3.818\pm 0.021$  \\
  12965.5343   &  $4.463\pm 0.022$ & 13355.4852   &  $3.397\pm 0.019$  \\
  12966.5492   &  $4.309\pm 0.027$ & 13356.4777   &  $3.397\pm 0.022$  \\
  12967.5156   &  $4.435\pm 0.026$ & 13357.4334   &  $3.360\pm 0.018$  \\
  12968.5385   &  $4.372\pm 0.024$ & 13358.4690   &  $3.454\pm 0.021$  \\
  12973.4510   &  $4.025\pm 0.132$ & 13365.5680   &  $3.342\pm 0.037$  \\
  12974.5110   &  $3.982\pm 0.131$ & 13379.5508   &  $3.316\pm 0.019$  \\
  12983.6310   &  $4.241\pm 0.045$ & 13383.4821   &  $3.446\pm 0.019$  \\
\hline
\end{tabular}
\end{table}

\begin{table}
  \contcaption{Combined F5170 continuum fluxes from spectral and photometric observations.}
  \begin{tabular}{@{}lclc}
  \hline
JD-2,440,000 & $F5170$  &Julian Date & $F5170$                   \\
\hline
  12984.5580   &  $4.277\pm 0.034$ & 13384.4410   &  $3.389\pm 0.020$  \\
  12985.4810   &  $4.329\pm 0.087$ & 13410.4190   &  $3.370\pm 0.113$  \\
  12988.5170   &  $4.372\pm 0.029$ & 13411.5060   &  $3.259\pm 0.106$  \\
  12996.5188   &  $4.248\pm 0.024$ & 13412.3370   &  $2.904\pm 0.097$  \\
  12997.4890   &  $4.193\pm 0.025$ & 13419.3710   &  $3.446\pm 0.047$  \\
  13423.3570   &  $3.454\pm 0.033$ & 13822.3430   &  $6.565\pm 0.043$  \\ 
  13424.4630   &  $3.439\pm 0.068$ & 13823.3110   &  $6.558\pm 0.044$  \\
  13425.3360   &  $3.493\pm 0.041$ & 13830.2380   &  $6.715\pm 0.113$  \\
  13434.3090   &  $3.703\pm 0.022$ & 13837.3880   &  $6.661\pm 0.051$  \\
  13436.3070   &  $3.791\pm 0.024$ & 13839.3400   &  $6.759\pm 0.068$  \\
  13437.2730   &  $3.483\pm 0.114$ & 13844.3300   &  $6.879\pm 0.040$  \\
  13441.3348   &  $3.908\pm 0.021$ & 13845.4410   &  $6.959\pm 0.047$  \\
  13445.2997   &  $4.160\pm 0.024$ & 13849.2870   &  $7.053\pm 0.251$  \\
  13446.3470   &  $4.388\pm 0.042$ & 13850.2917   &  $7.117\pm 0.041$  \\
  13449.3550   &  $4.369\pm 0.034$ & 13854.2850   &  $7.053\pm 0.220$  \\
  13459.3238   &  $4.875\pm 0.026$ & 13875.3320   &  $7.178\pm 0.086$  \\
  13460.2620   &  $4.978\pm 0.028$ & 13880.3050   &  $6.746\pm 0.036$  \\
  13461.2693   &  $5.067\pm 0.027$ & 13881.2850   &  $6.795\pm 0.041$  \\
  13462.3190   &  $5.040\pm 0.032$ & 13953.5550   &  $6.732\pm 0.106$  \\
  13464.2240   &  $5.285\pm 0.037$ & 13959.5530   &  $6.737\pm 0.080$  \\
  13465.3125   &  $5.363\pm 0.031$ & 13967.5620   &  $6.846\pm 0.056$  \\
  13471.2810   &  $5.882\pm 0.206$ & 13973.5680   &  $6.731\pm 0.050$  \\
  13476.3240   &  $5.693\pm 0.037$ & 13986.5780   &  $7.163\pm 0.090$  \\
  13478.2290   &  $5.844\pm 0.111$ & 13987.4830   &  $7.138\pm 0.066$  \\
  13487.2960   &  $5.574\pm 0.029$ & 13989.5540   &  $7.280\pm 0.091$  \\
  13493.2990   &  $5.485\pm 0.029$ & 13991.5330   &  $7.168\pm 0.069$  \\
  13495.2680   &  $5.497\pm 0.034$ & 13994.5660   &  $7.272\pm 0.080$  \\
  13508.3130   &  $5.755\pm 0.037$ & 13995.5930   &  $7.363\pm 0.054$  \\
  13509.2980   &  $5.987\pm 0.045$ & 14010.5960   &  $7.574\pm 0.066$  \\
  13611.5630   &  $6.791\pm 0.081$ & 14013.6130   &  $7.494\pm 0.089$  \\
  13612.5531   &  $6.801\pm 0.083$ & 14022.5790   &  $7.319\pm 0.058$  \\
  13621.5230   &  $6.877\pm 0.082$ & 14023.4520   &  $7.557\pm 0.061$  \\
  13641.5660   &  $6.279\pm 0.096$ & 14038.5550   &  $7.741\pm 0.089$  \\
  13644.5290   &  $5.884\pm 0.039$ & 14044.6030   &  $8.011\pm 0.123$  \\
  13645.6130   &  $5.900\pm 0.076$ & 14048.5820   &  $7.691\pm 0.097$  \\
  13647.5700   &  $6.158\pm 0.095$ & 14059.5450   &  $7.104\pm 0.047$  \\
  13648.6094   &  $5.937\pm 0.048$ & 14060.6550   &  $6.988\pm 0.147$  \\
  13649.5913   &  $6.174\pm 0.075$ & 14062.5450   &  $7.171\pm 0.054$  \\
  13650.6170   &  $5.939\pm 0.054$ & 14064.5904   &  $7.159\pm 0.049$  \\
  13651.5770   &  $5.914\pm 0.040$ & 14065.4730   &  $7.052\pm 0.036$  \\
  13653.6120   &  $5.885\pm 0.031$ & 14067.5428   &  $6.865\pm 0.045$  \\
  13654.5950   &  $5.972\pm 0.038$ & 14069.6400   &  $6.920\pm 0.056$  \\
  13670.5550   &  $6.284\pm 0.083$ & 14076.4610   &  $6.903\pm 0.156$  \\
  13676.5860   &  $6.309\pm 0.079$ & 14078.5510   &  $6.923\pm 0.101$  \\
  13680.5980   &  $5.868\pm 0.033$ & 14086.5590   &  $7.182\pm 0.097$  \\
  13683.5360   &  $5.688\pm 0.189$ & 14091.5530   &  $7.180\pm 0.044$  \\
  13698.4690   &  $6.366\pm 0.103$ & 14106.5310   &  $7.897\pm 0.243$  \\
  13702.4780   &  $5.872\pm 0.203$ & 14111.5780   &  $7.450\pm 0.050$  \\
  13708.5590   &  $6.684\pm 0.036$ & 14116.4500   &  $7.638\pm 0.059$  \\
  13724.5730   &  $7.009\pm 0.049$ & 14117.5520   &  $7.552\pm 0.331$  \\
  13728.3990   &  $6.749\pm 0.217$ & 14118.4140   &  $7.587\pm 0.055$  \\
  13733.5160   &  $6.918\pm 0.039$ & 14119.4440   &  $7.704\pm 0.051$  \\
  13738.4460   &  $6.940\pm 0.046$ & 14121.5630   &  $7.700\pm 0.047$  \\
  13739.5320   &  $6.813\pm 0.047$ & 14123.4870   &  $7.720\pm 0.064$  \\
  13744.5540   &  $6.880\pm 0.047$ & 14142.5940   &  $7.459\pm 0.097$  \\
  13747.4950   &  $6.735\pm 0.078$ & 14145.3596   &  $7.674\pm 0.049$  \\
  13749.4240   &  $6.698\pm 0.070$ & 14146.4190   &  $7.550\pm 0.044$  \\
  13760.3710   &  $6.082\pm 0.030$ & 14149.4380   &  $7.735\pm 0.041$  \\
  13761.4160   &  $6.143\pm 0.042$ & 14150.4260   &  $7.695\pm 0.038$  \\
  13763.3990   &  $5.901\pm 0.035$ & 14156.2700   &  $7.366\pm 0.132$  \\
  13774.4530   &  $5.677\pm 0.189$ & 14167.3450   &  $7.289\pm 0.050$  \\
  13787.3870   &  $6.090\pm 0.203$ & 14169.2930   &  $7.252\pm 0.048$  \\
\hline
\end{tabular}
\end{table}

\begin{table}
  \contcaption{Combined F5170 continuum fluxes from spectral and photometric observations.}
  \begin{tabular}{@{}lclc}
  \hline
JD-2,440,000 & $F5170$  &Julian Date & $F5170$                   \\
\hline
  13788.3440   &  $5.997\pm 0.091$ & 14171.3070   &  $7.250\pm 0.050$  \\
  13790.2820   &  $6.259\pm 0.199$ & 14174.4000   &  $7.481\pm 0.243$  \\
  13799.2870   &  $6.168\pm 0.046$ & 14180.3510   &  $7.516\pm 0.260$  \\
  13807.3650   &  $6.250\pm 0.057$ & 14181.2360   &  $7.412\pm 0.042$  \\
  13816.4150   &  $6.349\pm 0.041$ & 14191.2710   &  $7.835\pm 0.083$  \\
  13820.3430   &  $6.248\pm 0.208$ & 14200.3340   &  $8.249\pm 0.058$  \\
  14201.2850   &  $7.755\pm 0.269$ & 14523.3480   &  $6.185\pm 0.197$  \\
  14204.2350   &  $8.256\pm 0.058$ & 14530.3709   &  $6.735\pm 0.064$  \\
  14206.3000   &  $8.379\pm 0.073$ & 14532.3500   &  $6.735\pm 0.044$  \\
  14213.2460   &  $7.978\pm 0.049$ & 14534.3240   &  $6.662\pm 0.036$  \\
  14220.2400   &  $8.159\pm 0.108$ & 14535.3485   &  $6.570\pm 0.031$  \\
  14234.2640   &  $8.173\pm 0.060$ & 14536.3470   &  $6.542\pm 0.038$  \\
  14281.2840   &  $8.751\pm 0.203$ & 14537.3175   &  $6.452\pm 0.051$  \\
  14283.2890   &  $8.724\pm 0.117$ & 14538.3130   &  $6.436\pm 0.062$  \\
  14331.5430   &  $8.416\pm 0.086$ & 14542.2750   &  $6.073\pm 0.056$  \\
  14337.5760   &  $8.331\pm 0.118$ & 14554.3231   &  $6.021\pm 0.030$  \\
  14338.5040   &  $8.265\pm 0.051$ & 14555.2970   &  $5.994\pm 0.093$  \\
  14369.5160   &  $7.978\pm 0.124$ & 14565.4980   &  $5.939\pm 0.037$  \\
  14370.5390   &  $7.390\pm 0.197$ & 14567.2910   &  $5.906\pm 0.034$  \\
  14371.4990   &  $7.464\pm 0.073$ & 14568.2830   &  $6.029\pm 0.043$  \\
  14372.4860   &  $7.718\pm 0.071$ & 14574.2652   &  $6.262\pm 0.052$  \\
  14376.5510   &  $7.373\pm 0.061$ & 14582.4170   &  $6.398\pm 0.035$  \\
  14377.5820   &  $7.570\pm 0.110$ & 14585.2780   &  $6.093\pm 0.046$  \\
  14386.4570   &  $7.490\pm 0.098$ & 14586.3090   &  $6.284\pm 0.086$  \\
  14391.6032   &  $7.187\pm 0.051$ & 14587.2927   &  $6.252\pm 0.034$  \\
  14392.5390   &  $7.115\pm 0.097$ & 14588.2802   &  $6.184\pm 0.054$  \\
  14393.5960   &  $7.037\pm 0.045$ & 14590.3090   &  $6.278\pm 0.098$  \\
  14423.5722   &  $6.184\pm 0.037$ & 14591.2730   &  $6.440\pm 0.108$  \\
  14425.6040   &  $6.198\pm 0.032$ & 14596.3030   &  $6.331\pm 0.101$  \\
  14426.5370   &  $6.223\pm 0.047$ & 14600.2710   &  $6.832\pm 0.055$  \\
  14428.4230   &  $6.298\pm 0.058$ & 14601.3560   &  $6.695\pm 0.093$  \\
  14434.4860   &  $6.146\pm 0.044$ & 14602.3840   &  $6.727\pm 0.057$  \\
  14438.4380   &  $5.934\pm 0.100$ & 14603.3040   &  $6.667\pm 0.051$  \\
  14439.6060   &  $6.055\pm 0.034$ & 14604.3070   &  $6.752\pm 0.056$  \\
  14443.5395   &  $6.118\pm 0.031$ & 14617.2919   &  $6.978\pm 0.059$  \\
  14444.5270   &  $6.180\pm 0.033$ & 14618.2730   &  $6.876\pm 0.087$  \\
  14465.4540   &  $7.011\pm 0.042$ & 14620.3010   &  $7.053\pm 0.091$  \\
  14467.5530   &  $7.210\pm 0.043$ & 14622.4260   &  $7.140\pm 0.087$  \\
  14472.4660   &  $7.411\pm 0.041$ & 14628.3120   &  $6.876\pm 0.046$  \\
  14475.4050   &  $6.727\pm 0.227$ & 14632.3170   &  $6.960\pm 0.075$  \\
  14476.4683   &  $7.358\pm 0.048$ & 14643.3460   &  $7.117\pm 0.044$  \\
  14477.4660   &  $7.463\pm 0.058$ & 14647.3060   &  $7.129\pm 0.041$  \\
  14478.4300   &  $7.452\pm 0.053$ & 14687.5470   &  $6.498\pm 0.095$  \\
  14479.4230   &  $7.456\pm 0.053$ & 14713.5470   &  $7.087\pm 0.267$  \\
  14480.4251   &  $7.368\pm 0.049$ & 14718.5400   &  $7.288\pm 0.051$  \\
  14481.4803   &  $7.412\pm 0.054$ & 14720.5630   &  $7.213\pm 0.227$  \\
  14482.4074   &  $7.451\pm 0.053$ & 14738.5060   &  $7.030\pm 0.237$  \\
  14483.4110   &  $7.452\pm 0.069$ & 14740.4520   &  $7.053\pm 0.241$  \\
  14484.3690   &  $7.183\pm 0.069$ & 14742.5900   &  $7.451\pm 0.051$  \\
  14488.4530   &  $7.111\pm 0.120$ & 14769.5830   &  $8.158\pm 0.265$  \\
  14490.4630   &  $7.314\pm 0.062$ & 14778.5960   &  $8.054\pm 0.050$  \\
  14496.4231   &  $7.715\pm 0.044$ & 14779.6064   &  $8.005\pm 0.041$  \\
  14497.4230   &  $7.654\pm 0.054$ & 14780.5790   &  $8.052\pm 0.042$  \\
  14498.4680   &  $7.550\pm 0.051$ & 14781.5400   &  $7.969\pm 0.048$  \\
  14499.4361   &  $7.641\pm 0.045$ & 14784.3780   &  $8.013\pm 0.128$  \\
  14500.4060   &  $7.739\pm 0.050$ & 14786.4900   &  $8.206\pm 0.084$  \\
  14502.4057   &  $7.653\pm 0.052$ & 14796.4990   &  $8.174\pm 0.050$  \\
  14503.4437   &  $7.605\pm 0.045$ & 14799.6310   &  $8.263\pm 0.061$  \\
  14508.4223   &  $7.310\pm 0.034$ & 14801.3500   &  $8.467\pm 0.064$  \\
  14512.5400   &  $6.981\pm 0.131$ & 14802.5980   &  $8.346\pm 0.052$  \\
  14522.2720   &  $6.496\pm 0.211$ & 14804.5660   &  $8.214\pm 0.093$  \\
\hline
\end{tabular}
{\\\footnotesize 
Continuum fluxes are in units $10^{-15}$\,ergs\,cm$^{-2}$\,s$^{-1}$\,\AA$^{-1}$  
}
\end{table}

Bottom panel in Figure~\ref{lc} shows the combined light curve from different telescopes.
The combined light curve shows long time scale continuum variability in Mrk~6 as well as more rapid random changes. The flux maxima were observed in 1995--1996 and in 2007.

Statistical parameters of the light curves for the lines and continuum are listed in Table~\ref{stat}. Column~1 gives the spectral features; column~2 gives a number of data points; column~3 is the median time interval between the data points. The combined continuum light curve is sampled better than the \hb\ light curve, the \hb\ light curve is sampled better than \ha. The mean flux and standard deviation are given in columns~4 and 5, and column~6 lists the variance $F_{var}$ calculated as the ratio of the rms fluctuation, corrected for the effect of measurement errors, to the mean flux. $R_{max}$ in column~7 is the ratio between the maximum and minimum fluxes corrected for the measurement errors. Uncertainties in $F_{var}$ and $R_{max}$ were computed assuming that a light curve is a set of statistically dependent values, i.e., a random process. The $F_{var}$ values in Table~\ref{stat} are the lowest limits of actual $F_{var}$ because the observed fluxes were not corrected for starlight contamination.

\begin{table}
\caption{Light curve statistics.}
\label{stat}
{\scriptsize
\begin{tabular}{@{}ccccccc@{}}
 \hline
Time   &  N & dt-med & Mean    & std$^b$  & $F_{var}$& $R_{max}$ \\
series$^a$ &    &(days)& Flux$^b$&          &          &           \\
(1)    &(2) & (3)    &(4)      &  (5)     & (6)      & (7)       \\
\hline
\multicolumn{7}{@{}c}{1992--2008, JD2448630--2454804}\\
F5170s   & 235 &14 & 6.093 & 1.101& 0.18$\pm$0.03 & 2.65$\pm$0.06 \\
F5170scp & 742 & 3 & 5.780 & 1.277& 0.22$\pm$0.05 & 3.16$\pm$0.08 \\
\hb      & 235 &14 & 4.065 & 0.658& 0.16$\pm$0.03 & 2.70$\pm$0.09 \\[3pt]

F7030s   & 102 &29 & 6.897 & 1.261& 0.18$\pm$0.04 & 2.43$\pm$0.17 \\
\ha      & 102 &29 & 32.06 & 3.77 & 0.11$\pm$0.02 & 1.83$\pm$0.09 \\[3pt]
\hline
\end{tabular}
} {\\\footnotesize $^a$ The letters in the first column indicate the origin of the continuum fluxes:
"s" -- from the spectral observations only and "scp" -- from combined spectral and photometric
observations ("c" -- CCD photometry, "p" -- photoelectric photometry).\\ $^b$ Mean fluxes and
standard deviations (std) are in units of
10$^{-13}$~erg~cm$^{-2}$~s$^{-1}$ and 10$^{-15}$~erg~cm$^{-2}$~s$^{-1}$~\AA$^{-1}$ for the
lines and the continuum, respectively.}
\end{table}

\section{Cross-correlation between the continuum and the integral Balmer line flux
variations}
\label{fccf}
As mentioned in the Introduction, estimating the light travel time delay between the continuum
and emission line flux variations is of special relevance for the determination of the BLR size,
which, in turn, can be used for black hole mass measurements \citep[see][]{WPM99, Peterson04}.
This time delay (or lag) is estimated through the cross-correlation function (CCF).
\citet{KorGask91} demonstrated that the CCF centroid gives the luminosity-weighted radius, in
contrast to the CCF peak, which is more influenced by gas at small radii, according to
\citet{Gaskell86}.

The time delays were computed using the interpolated cross-correlation function (ICCF)
\citep{Gaskell86, White94, Peterson01}. We computed both the lag related to the CCF peak ($\tau_{pk}$) and the CCF centroid ($\tau_{cn}$). The CCF centroid was adopted to be measured above the correlation level at $r\geq0.8 r_{max}$. The lag uncertainties were computed using the model-independent Monte Carlo flux randomization/random subset selection (FR/RSS) technique described by \citet{Peterson98}. The number of realizations was as large as 4000. The uncertainties were computed from the distribution function for $\tau_{pk}$ and $\tau_{cn}$ at the 68\% confidence, which corresponds to $\pm 1\sigma$ errors for the normal distribution.

The spectra for the 1992 season were obtained with the 2$\arcsec$ entrance slit and they were discarded from the CCF analysis in order to exclude the aperture effects. The results of the cross-correlation analysis for the 1993--2008 interval are presented in Table~\ref{ccf-ab}. The meaning of symbols "s" and "scp" in the first and second columns of Table~\ref{ccf-ab} is the same as in Table~\ref{stat}: "s" -- from spectral observations only and "scp" -- from combined spectral and photometric observations ("c" -- CCD, "p" -- photoelectric). 

\begin{figure}
\includegraphics[width=84mm]{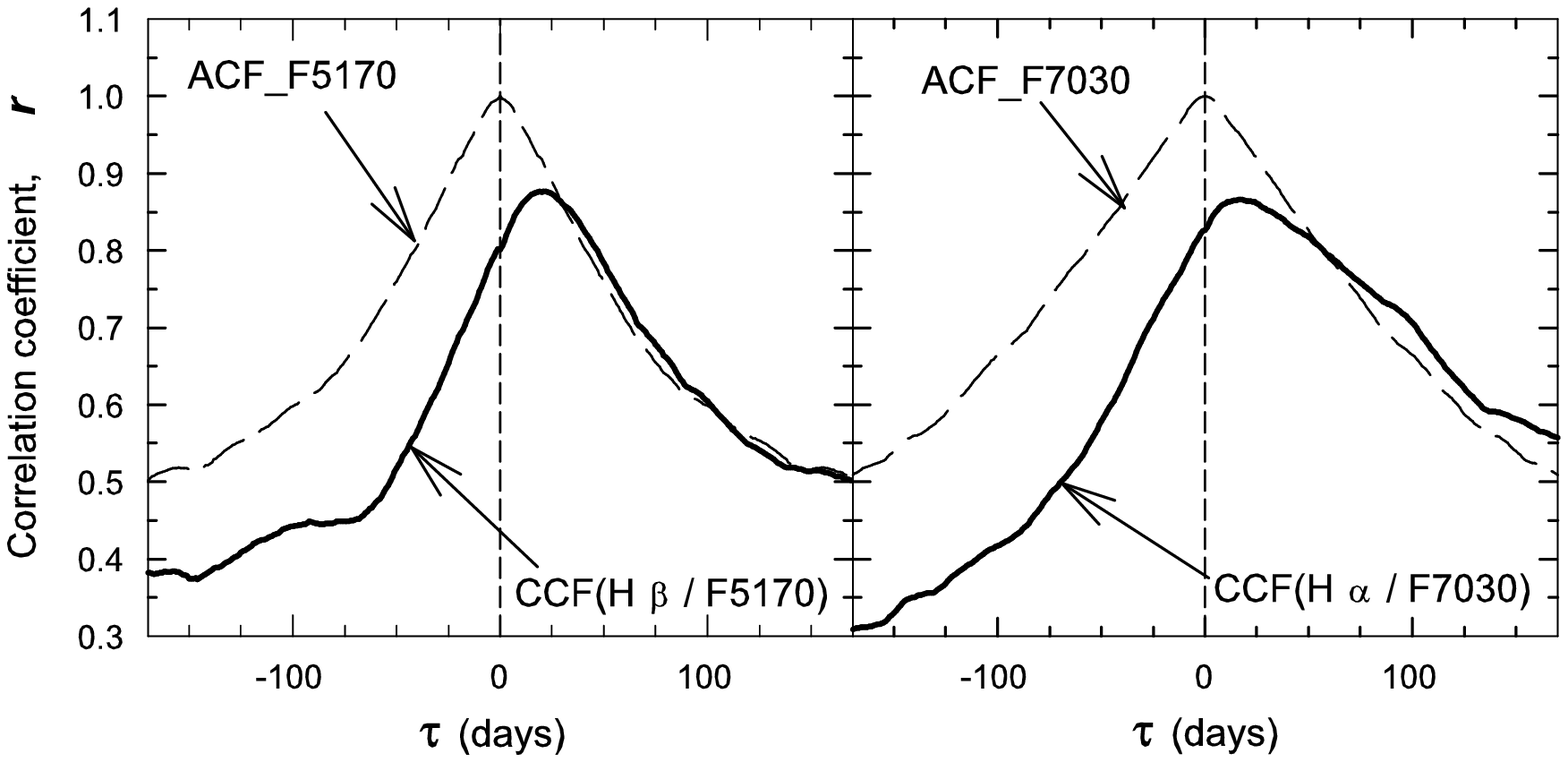}
\caption{CCFs of Mrk~6  between the \hb\ and continuum F5170 fluxes (left panel) as well
as between the \ha\ and continuum F7090 in 1993--2008 (thick lines). ACFs are shown by dashed
lines.}
\label{ccf}
\end{figure}

\begin{table}
\caption{Cross-correlation results for 1993--2008.}
\label{ccf-ab}
{\footnotesize
\begin{tabular}{@{}cccccc}
 \hline
First set & Second set &$ r_{max}$&$\tau_{cn}$ & $\tau_{pk}$\\
\hline

\hb    & F5170s   & 0.84 & 21.5$^{+3.9}_{-2.1}$  & 21.0$^{+1.6}_{-4.9}$\\[5pt]
\hb    & F5170scp & 0.83 & 21.9$^{+2.6}_{-2.7}$  & 21.0$^{+1.2}_{-4.0}$\\[5pt]
\ha    & F7030s   & 0.88 & 31.6$^{+10.9}_{-8.4}$ & 18.0$^{+8.3}_{-6.4}$\\[5pt]
\ha    & F5170scp & 0.85 & 26.8$^{+10.6}_{-4.8}$ & 22.0$^{+6.9}_{-6.9}$\\[5pt]
\ha    & \hb      & 0.93 & 8.5$^{+8.6}_{-4.2}$   &  2.5$^{+0.4}_{-1.6}$\\[5pt]
F7030s & F5170s   & 0.97 & 7.2$^{+3.7}_{-4.4}$   &  5.5$^{+2.5}_{-1.5}$\\[5pt]
F7030s & F5170scp & 0.96 & 8.4$^{+2.1}_{-3.6}$   &  3.2$^{+4.4}_{-1.8}$\\[5pt]
\hg    & F5170sc  & 0.53 & 33.1$^{+9.2}_{-11.6}$ & 27.1$^{+17.1}_{-5.9}$\\[5pt]
\hg    &\hb       & 0.82 &  8.9$^{+8.1}_{-9.1}$  & $-0.6^{+18.3}_{-2.3}$&\\[5pt]
\hline
\end{tabular}
}
{\\\footnotesize
Meaning of symbols "s"\ and "scp"\ the same as in Table\ref{stat}.}
\end{table}

Figure~\ref{ccf} shows the cross-correlation results for the \hb\, and \ha\, line fluxes with
the continuum as well as autocorrelation functions of the continuum for the 1993--2008 time
interval. Table~\ref{ccf-ab} gives the cross-correlation results. When the continuum light curve
is a combination of the spectral and the photometric data (designated as $F5170scp$ in Table~\ref{ccf-ab}) then the CCF computation has been carried out with the rebinning the $F5170scp$ light curve to the times of observations of the first time series. This is because the combined continuum light curve has much more data points than the continuum light curves which consist from the spectral data only (designated as $F5170s$ in Table~\ref{ccf-ab}). For the $F5170s$ light curve, we follow a standard method of the CCF computation with rebinnig both time series.

The variations of the \ha\ and \hb\ fluxes are tightly correlated as well as variations of the continuum fluxes near both lines (the correlation coefficients are equal to 0.93 and 0.97, respectively). The positive lag values for the `\ha--\hb' and `F7030--F5170' light curves
mean that the region of effective continuum emission at $\lambda$5170 is probably smaller than that at $\lambda$7030 and the region of effective \hb\ emission is probably smaller than that \ha. The probability that the delay for the `F7030--F5170' and `\ha--\hb' is less than zero equals to 0.023 and 0.036, respectively.

The time interval 1993--2008, that has been used for cross-correlation analysis, is very long.
We have divided it into five subinterval in order to check whether the lag values for individual sub-intervals are the same as for the entire 1993--2008 period, whether they are changed in time, and whether they are correlated with the fluxes and with line widths. In particular, the effective region of the broad-line emission can depend on the incident continuum flux, so the lag can depend on the continuum flux. Under a virialized motion of the line-emitting gas the expected relation between the lag $\tau$ and the line width $V$ is as follows: $V \propto \tau^{-1/2}$. To this end, we used only the \hb\ spectra because more reliable lag estimate for this line, and we carried out the cross-correlation analysis separately for each of the five time intervals listed in the Table \ref{time intervals}. The first and second intervals were taken the same as in the paper by \citet{Ser99}.

\begin{table}
\caption{Time intervals for time series analysis in the \hb\ region.}
\label{time intervals}
{\small
\begin{tabular}{@{}lccccc}
\hline
Time   &JD2440000+ &\multicolumn{2}{c}{N}& \multicolumn{2}{c}{$\Delta t_{med}$} \\
Series &           & \hb & cont$^a$ &\hb &cont \\
\hline
1& 09250--09872 & 38 & 51\,sp  & 14.0 & 8.9\\
2& 09980--10777 & 54 & 90\,sp  & 11.9 & 3.0\\
3& 10869--11516 & 27 & 59\,sp  & 16.0 & 6.1\\
4& 11557--13356 & 47 &266\,scp & 19.1 & 2.0\\
5& 13611--14804 & 58 &242\,scp & 10.8 & 2.1\\
\hline
\end{tabular}
}
{\\\footnotesize $^a$ Meaning of symbols "sp"\ and "scp"\ is the same as in Table\ref{stat}.\\}
\end{table}

The cross-correlation results for the five time subintervals are given in Table~\ref{ccf-b}.
For the continuum light curve in Table~\ref{ccf-b} that consists of the spectral data only
(i.e., when the number of data points in both the \hb\ and continuum light curves is the same)
we have used the standard method of the CCF computation with the rebinning both time series.
When the photometric data are added to the continuum light curve, only the continuum light curve
was rebinned to the times of observation of the \hb\ light curve. As for the entire period, the lag uncertainties for the individual subintervals were computed by using the FR/RSS method as given by \citet{Peterson98}. The uncertainties for each subinterval were computed using 4000 FR/RSS realizations, and the probability distributions for both $\tau_{cn}$ and $\tau_{pk}$ were calculated. Each of these distributions was found to be very different from the normal distribution. 
{\bf To obtain the unweighted mean lag and its uncertainties we 
have sequentially (one after another) performed the convolution of the five individual distributions and then we have scaled the $\tau$-axis by dividing it by a number of subintervals (i.e., by five). After the convolution operation, the final probability distribution was found to be almost normal with the expectation and standard deviation as given in Table~\ref{ccf-b} for the mean lag.} 

\begin{table}
\caption{Cross-correlation results for the \hb\ line for the five subintervals.}
\label{ccf-b}
{\footnotesize
\begin{tabular}{@{}lccccc}
 \hline
 Subset&\multicolumn{2}{c}{N points}&$\tau_{cn}$& $\tau_{pk}$ & $r_{max}$\\
Number&  \hb & F5170   &                &        &        \\
\hline
1 & 38 & 38     &  22.7$^{+7.4}_{-3.0}$  &20.5$^{+4.7}_{-2.8}$  & 0.939\\[5pt]
2 & 54 & 54     &  20.8$^{+3.0}_{-2.6}$  &18.1$^{+3.6}_{-1.0}$  & 0.942\\[5pt]
3 & 27 & 27     &  19.3$^{+4.0}_{-6.0}$  &18.2$^{+7.7}_{-6.1}$  & 0.761\\[5pt]
4 & 47 &199$^a$ &  26.2$^{+12.3}_{-6.8}$ &14.2$^{+14.6}_{-1.4}$  & 0.898\\[5pt]
5 &58  &230$^a$ &  20.2$^{+5.0}_{-3.9}$  &26.8$^{+0.3}_{-13.8}$  & 0.794\\[5pt]
\hline
\multicolumn{3}{l}{Mean value:}& 21.4$\pm$2.0& 19.3$\pm$1.9\\
\hline
1 & 38 & 51$^b$ & 21.2$^{+4.0}_{-3.2}$   &21.0$^{+4.0}_{-3.4}$  & 0.923\\[5pt]
2 & 54 & 90$^b$ & 20.7$^{+3.0}_{-2.4}$   &22.0$^{+1.3}_{-4.7}$  & 0.939\\[5pt]
3 & 27 & 59$^b$ & 20.5$^{+5.6}_{-7.0}$   &28.5$^{+1.4}_{-21.9}$ & 0.683\\[5pt]
4 & 47 &266$^c$ & 23.9$^{+17.0}_{-7.3}$  &9.5$^{+15.6}_{-1.2}$  & 0.881\\[5pt]
5 & 58 &242$^c$ & 20.4$^{+4.6}_{-4.1}$   &21.8$^{+5.1}_{-8.7}$  & 0.803\\[5pt]
\hline
\multicolumn{3}{l}{Mean value}& 21.1$\pm$1.9   & 20.5$\pm$2.2\\
\hline
\end{tabular}
}
{\\\footnotesize
$^a$ Continuum light curve was combined from the spectral data and from the CCD photometry.\\
$^b$ Continuum light curve was combined from the spectral data and from the photoelectric photometry.\\
$^c$ Continuum light curve was combined from the spectral data and from both the CCD and photoelectric photometry.\\
}
\end{table}

Since there are large gaps in our time series, we decided to investigate in more detail their effect on the lag uncertainties for Mrk~6. To make sure that the uncertainties in our lag estimates are realistic, we decided to verify the effect of the sampling of our time series to the lag determination and to compare results with those obtained by random subset selection (RSS) method. We have generated random time series with the same autocorrelation function as observed ACF. Stationary random process (or time series) with a given ACF can be generated from an array of {\em independent} random values $\xi_1, \xi_2, \xi_3, \ldots, \xi_n $. To do this, it is necessary to find a matrix $U_{ij}$, such that the multiplication of the matrix $U$ to the vector $\xi$ gives a vector of {\em dependent} random values $x_1, x_2, x_3, \ldots, x_n$, with a given correlation matrix $\rho_{ij} = ACF(t_j - t_i)$, where $t_1, t_2, t_3, \ldots, t_n$ are times of observations. The matrix $U$ is related to the correlation matrix $\rho$ as follows: 
\begin {equation} 
  \rho = U\, U ^{\rm T}, 
\label{genrandproc} 
\end {equation} 
where $T$ denotes a transposition operation. We used our own algorithm to compute the matrix $U$ from ACF.

First we have generated 1000 realizations of the continuum light curve with a time resolution of 1 day over a period of 6452 days, which is longer than the real observed time interval (6175 days). 

To simulate the \hb\ light curve from the continuum light curve we have experimented with the three kinds of transfer functions: (1) delta-function $\delta(\tau - 20)$, (2) $\Pi$-shaped function which is a constant for $0<\tau<40$, and (3) triangular function which is linearly decreased down to a zero value from $\tau = 0$ to $\tau=68.3$. Here the lag $\tau$ are in units of days. After the convolution with the simulated continuum light curves, all the transfer functions give the \hb\ light curve with a lag of about 20 days. Next the simulated continuum and line light curves were rebinned to real moments of observations and the cross-correlation functions were computed for each realization. The lag peak and centroid were measured. The largest uncertainties were obtained when we used the triangular transfer function. These uncertainties are only due to the sampling of the observation data. Table~\ref{simul} gives a comparison of both methods for uncertainty estimates (i.e., the random time series versus RSS method) for the triangular transfer function.
The columns designated as $\tau_{pk}$ and $\tau_{cn}$ are a lag with $\pm 1\sigma$ uncertainties computed from the random time series, while other two columns designated as RSS give $\pm 1\sigma$ uncertainties computed from RSS method for $\tau_{pk}$ and $\tau_{cn}$, respectively. The last column of the table is the mean CCF peak value obtained from the method of the random time series. As can be seen from this table, the RSS method gives comparable or larger (up to two times) uncertainties than the method of the random time series does. The random time series method seems to be more direct way to estimate lag uncertainties and since the lag uncertainties were computed in this paper by the RSS method, they seem to be realistic or slightly overestimated. 

There is a contradiction between our lag estimate and the preliminary results on Mrk~6 published in a conference proceeding by Grier et al. (NASA ADS tag 2011nlsg.confE..52G). This new campaign had a nightly sampling rate and it spanned 125 nights beginning 2010 August 31 and ending on 2011 January 3. They claimed a lag of $8\pm 3$ days for the \hb\ line in Mrk 6. We have first checked the effect of the removing of linear trends from our light curves {\bf \citep[as recommended by some authors, e.g.,][]{Welsh}}. We have removed linear trends from the \hb\ and continuum light curves for each subinterval, even if there are no such trends exist. Then we have recalculated a mean lag value and it was to be $19$ days, i.e. two days less. Then we have generated random time series (continuum and \hb) by exactly the same way as described above, but for the sampling rate of Grier et al. in order to check whether the duration of the monitoring program is important for the lag measurements. We found that with the data sampling of Grier et al. our lag estimate must be less by one more day, and so a total difference between our and their results must be three days. The real difference is much larger than three days. An unexpected result of our simulation was very large lag uncertainties for the 125-days data sampling and for the triangular transfer function for the \hb\ line (see above). The lag uncertainties were found to be as large as $\pm 6$ days! However, for the $\delta(\tau-20)$ transfer function (i.e., when the line light curve is simply a shifted version of the continuum light curve) the lag uncertainties were found to be as small as $\sim 0.1$ days. So, for short-term campaigns and for transfer functions with a long tail (i.e., when the line light curve is not only shifted, but a strongly smoothed version of the continuum light curve), the uncertainties in lag estimates can be very large and they are connected to the extrapolation of the line and continuum fluxes when computing the CCF, not to the data sampling. We concluded that we can only explain the difference of three days between our and their lag measurements. Probably, the rest of the difference is due to real changes of lag or due to the measurement uncertainties and their underestimation.

It can be seen from Table~\ref{simul} that the simulated lag values are almost the same for all 
subinterval. The expected lag value can be obtained by convolving the ACF with the transfer function and it was found to be: $\tau_{pk} = 20.7$\,days and $\tau_{cn} = 22.4$\,day in an excellent agreement with the simulation results. So, the large gaps in our time series do not shift the lag measurements.

\begin{table}
\caption{Comparison of the lag uncertainty estimates between the method of the random time series and the RSS method.}
\label{simul}
{\small
\begin{tabular}{@{}ccccccc}
 \hline
Interval & $\tau_{pk}$& RSS & $\tau_{cn}$& RSS & $r_{sim}$\\
\hline
1& 20.8$^{+2.4}_{-2.1}$& 2.1& 22.2$^{+2.8}_{-2.8}$& 2.7& 0.968\\ [5pt]
2& 20.8$^{+1.5}_{-1.6}$& 1.6& 22.2$^{+2.8}_{-2.8}$& 2.1& 0.960\\[5pt]
3& 20.5$^{+2.8}_{-2.8}$& 6.0& 22.4$^{+4.6}_{-4.6}$& 4.3& 0.940\\[5pt]
4& 20.4$^{+2.6}_{-2.6}$& 3.4& 22.1$^{+4.4}_{-4.0}$& 8.5& 0.980\\[5pt]
5& 20.6$^{+2.4}_{-2.7}$& 6.9& 22.4$^{+3.7}_{-3.5}$& 3.3& 0.978\\[5pt]
\hline
\end{tabular}
}
\end{table}

\begin{table}
 \caption{The broad \hb, and \ha\ line width measurements for 1993--2008.}
 \label{width-lines}
{\small
\begin{tabular}{@{}lcc}
 \hline
        &FWHM & $\sigma_{line}$ \\
&\multicolumn{2}{c}{km\,s$^{-1}$} \\
\hline
\hb(mean-spectrum)& $6278 \pm 378 $&$ 2821 \pm 13 $\\
\hb(RMS-spectrum) & $5445 \pm 468 $&$ 2884 \pm 97 $\\
\hline \noalign{\smallskip}
\ha(mean-spectrum)& $5322 \pm 142 $&$ 2870 \pm 22 $\\
\ha(rms-spectrum) & $4443 \pm 338 $&$ 2780 \pm 35 $\\
\hline
\end{tabular}
}
\end{table}

\section{Line width measurements}
\label{linewidth}
It is well known that the emission-line profile evolution cannot be entirely attributed to the
reverberation effect and that the profile changes usually occur on a time scale that is much longer than the flux-variability time scale \citep[see][]{Wanders}. To decrease the effect of the long-term profile changes on the line width measurements and to get sufficient statistics, we measured the \hb\ and \ha\ line widths for the five subintervals presented in Table \ref{time intervals}. The line width is typically characterized by its full width at half maximum (FWHM) or by the second moment of the line profile, denoted as $\sigma_{line}$. To measure FWHM for the mean \ha\ and \hb\ line profiles, we removed narrow lines from the broad line profiles. This is not required for rms profiles. However, the spectra must be optimally aligned in wavelength and in spectral resolution in order to reduce the narrow line residuals in the rms profiles. It is difficult to measure FWHM because both the mean broad and rms \hb\ profiles are double-peaked, and so the scatter in the FWHM measurements is much larger than in $\sigma_{line}$ which is well defined for arbitrary line profiles. The uncertainties in the line width were obtained using bootstrap method described by \citet{Peterson04}. The \hb\ and \ha\ line width (both FWHM and $\sigma_{line}$) and their uncertainties are listed in Table~\ref{width-lines} for 1993--2008. Table~\ref{flx-ccf-width} gives the $\sigma_{line}$ computed separately for each considered subinterval and for the \hb\ line only. 

\begin{figure}
\includegraphics[width=84mm]{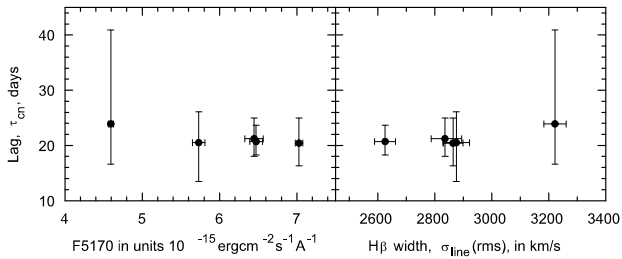}
\caption{The relation between the continuum flux and the lag ($\tau_{cn}$) for \hb\ in Mrk~6, as well as between the \hb\ width and the lag ($\tau_{cn}$).}
\label{lag-flx-wdt}
\end{figure}

We examined the relationship between the \hb\ lag, \hb\ width, and the continuum flux (see Fig.~\ref{lag-flx-wdt}). No significant correlations among above three parameters were found. In particular, the virial relationship between the lag and width does not contradict to our data, but the correlation coefficient between them does not differ significantly from the zero value.
More subintervals and less lag uncertainties are required.

\begin{table*}
\caption{Results for the five subintervals}
\label{flx-ccf-width}
{\small
\begin{tabular}{@{}lccccccc}
 \hline
Time &\multicolumn{2}{c}{Mean Flux$^a$} &$\tau_{cn}$ &\multicolumn{2}{c}{$\sigma_{line}$km s$^{-1}$} &\multicolumn{2}{c}{M$_{BH}^{b}$in units 10$^8$M$_{\odot}$}\\
series &F(5170)& F(\hb)&days &(mean)&(rms)&(mean)&(rms)\\
1&2&3&4&5&6&7&8\\
\hline
1& 6.446 &3.632 &21.2$^{+4.0}_{-3.2} $ &2813$\pm$13 &2836$\pm$48 &1.77$^{+0.33}_{-0.27}$ &1.80$^{+0.34}_{-0.28}$\\[5PT]
2& 6.472 &3.976 &20.7$^{+3.0}_{-2.4} $ &2804$\pm$ 6 &2626$\pm$37 &1.72$^{+0.25}_{-0.20}$ &1.50$^{+0.22}_{-0.18}$\\[5PT]
3& 5.730 &3.587 &20.5$^{+5.6}_{-7.0} $ &2808$\pm$14 &2876$\pm$46 &1.70$^{+0.46}_{-0.58}$ &1.79$^{+0.49}_{-0.61}$\\[5PT]
4& 4.594 &2.677 &23.9$^{+17.0}_{-7.3}$ &2870$\pm$13 &3222$\pm$39 &2.07$^{+1.50}_{-0.63}$ &2.62$^{+1.80}_{-0.80}$\\[5PT]
5& 7.027 &3.523 &20.4$^{+4.6}_{-4.1} $ &2807$\pm$ 8 &2864$\pm$35 &1.69$^{+0.38}_{-0.34}$ &1.77$^{+0.40}_{-0.36}$\\[5PT]
\hline
\multicolumn{3}{l}{Average:}&$21.1\pm1.9$ & $2812\pm 10^c$ & $2882\pm100^c$ & $1.76\pm0.16$& $1.85\pm0.21$\\
\hline
\end{tabular}
}
{\\\footnotesize
\mbox{\hspace{2.5cm}}$^a$ Same flux units as in Table 1 for 5170 \AA\ continuum and \hb, respectively. \hfill \mbox{~}\\
\mbox{\hspace{2.5cm}}$^b$ Using Onken et al (2004) calibration, $f=5.5$. \hfill \mbox{~}\\
{\bf \mbox{\hspace{2.5cm}}$^c$ The line width and uncertainties were computed as weighted average and assuming \hfill \mbox{~}\\
\mbox{\hspace{2.5cm}}$^{~}$ different expectations of the line widths among individual periods of observations.\hfill \mbox{~}\\}}
\end{table*}

\section{Black hole mass of Mrk~6}
\label{mass-bh}
Determination of the black hole mass from reverberation mapping rests upon the assumption that
the gravity of the central super-massive black hole dominates over gas motions in the BLR. The black
hole mass is defined by the virial equation $$ M_{BH}=f \frac{c\tau(\Delta V)^2}{G},$$
where $\tau$ is the measured emission-line time delay, $c$ is the speed of light, $c\tau$ represents
the BLR size, and $\Delta V$ is the BLR velocity dispersion. The dimensionless parameter $f$ is
the scaling factor, which depends on the BLR structure, kinematics and inclination of BLR. \citet{Peterson04}
argued that $\tau_{cn}$ for the time delay $\tau$, and $\sigma_{line}$, measured from the \hb\ emission line in the rms spectrum for the emission line width $\Delta V$, provide the most robust estimates of the black hole mass with the reverberation technique. Later \citet{Collin2006} confirmed that in most cases for the black hole mass estimate the line dispersion $\sigma_{line}$ is more suitable than the FWHM, and $\sigma_{line}$ from the rms-spectrum is more suitable than the $\sigma_{line}$ from the mean spectrum.

We adopt an average value of $f = 5.5$ based on the assumption that AGNs follow the same $M_{BH}-\sigma_*$ relationship as quiescent galaxies \citep{Onken04}. This is consistent with \citet{Woo10} and allows easy comparison with previous results, but this is about a factor of
two larger than the value of $f$ computed by \citet{Graham11}. The value of $f$ can be more decreased due to the effect of radiation pressure, as was explored by \citet{Marconi08, Marconi09}. Marconi et al. suggested that neglecting the effect of radiation pressure can lead to underestimation of the true black hole mass, especially in objects close to their Eddington limit. Discussion between \citet{Marconi08, Marconi09} and \citet{Netzer09} shows that there are many unclear questions in this area. Naturally, a corrective term for radiation pressure will decrease the $f$-factor.

We calculated the black hole mass for Mrk~6 with the use of $\tau_{cn}=21.1\pm 1.9$ for the
time delay averaged over five time intervals and $\sigma_{line}=2882\pm100$ km\,s$^{-1}$ from the rms spectra for \hb. With $\tau_{cn}$ taken in days and $(\Delta V)$ in km\,s$^{-1}$, and taking into account the time dilation correction for the value of $\tau_{cn}$, the mass is equal to: 
$$M_{BH}/M_{\odot}=0.1952\times f\times\frac{\tau_{cn}(obs)}{(1+z)}\times(\Delta V)^2.$$
The black hole mass calculated from the \hb\ line is $(1.85\pm0.21)\times 10^8$M$_{\odot}$. For the \ha\ line, the $\tau_{cn}=26.8\pm7.7$ days and $\sigma_{line}=2780\pm35$ km\,s$^{-1}$ and the black hole mass is equal to (2.2$\mathbf\pm0.6)\times10^8$M$_{\odot}$.

The black hole masses calculated for each of five periods of observations are listed in columns 7 and 8 of Table~\ref{flx-ccf-width} for the $\sigma_{line}$ from the mean and rms spectra. One can see that all the estimates of the black hole mass based on the \hb\ line are the same within the scatter.

\section{The BLR~Size--Luminosity and Mass--Luminosity Relationships}
\label{slm}
Many characteristics of the Mrk~6 galaxy are typical for active galaxies. In this connection, it
is of interest to see the localization of this galaxy on the BLR Radius--Luminosity and Mass--Luminosity diagrams. These diagrams determine a relationship between fundamental characteristics of AGNs. To this end, the Mrk~6 luminosity should be known. Up to the present, for many AGNs the luminosity in the rest-frame $\lambda_0=5100$ \AA\, has been corrected for host-galaxy starlight contribution within the apertures used in spectral observations \citep[see][]{Bentz09a}. These authors used high-resolution {\it Hubble Space Telescope} (HST) images to measure the starlight contribution. This contribution was found to be significant, especially for low-luminosity AGNs.

\begin{figure}
\includegraphics[width=70mm]{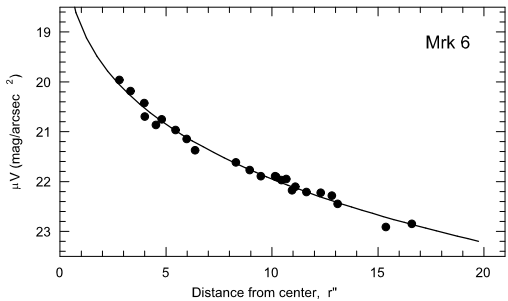}
\caption{Surface brightness distribution for Mrk~6 in the V-band on the basis of multi-aperture photoelectric photometry by \citet{Neizvestny}. The solid curve corresponds to the model $\mu (r)= a+br+cr^{1/4}$.}
\label{muV}
\end{figure}

We tried to get at least a rough estimate of the host-galaxy contribution using the observations made by \citet{Neizvestny} at the Special Astrophysical Observatory (SAO) in October 1984 with different apertures from A=4$\farcs$3 to 55$\arcsec$. The surface brightness distribution in the host galaxy of the Mrk~6 nucleus calculated on the basis of these measurements is shown in Fig.~\ref{muV}.

The galaxy contribution in our 3\arcsec$\times$11\arcsec\ spectral window was found to be $V_{gal}$=15\fm6 or $F_{gal}$=2.08$\times10^{-15}$erg\,s$^{-1}$cm$^{-2}$\AA$^{-1}$. The mean flux observed in the continuum near $\lambda_0$=5100~\AA\, is $F_{(gal+nuc)}$=6.093$\times10^{-15}$ergs$^{-1}$cm$^{-2}$\AA$^{-1}$ (see Table~\ref{stat}) and, thus, the mean flux corrected for the galaxy contribution is equal to $F_{nuc}$=4.013$\times10^{-15}$ergs$^{-1}$cm$^{-2}$\AA$^{-1}$.
The variability amplitude $F_{var}$ increases from 18\% to 27\% after accounting for the galaxy contribution. The mean flux was also corrected for Galactic reddening according to the NASA/IPAC Extragalactic Database (NED) \citet{Schlegel98}. The luminosity was found to be $\lambda L_{\lambda}$(5100)=(2.51$\pm0.78)\times10^{43}$~erg\,s$^{-1}$ adopting the galaxy distance D=81\,Mpc and when the galaxy contribution is removed.

\begin{figure}
\includegraphics[width=84mm]{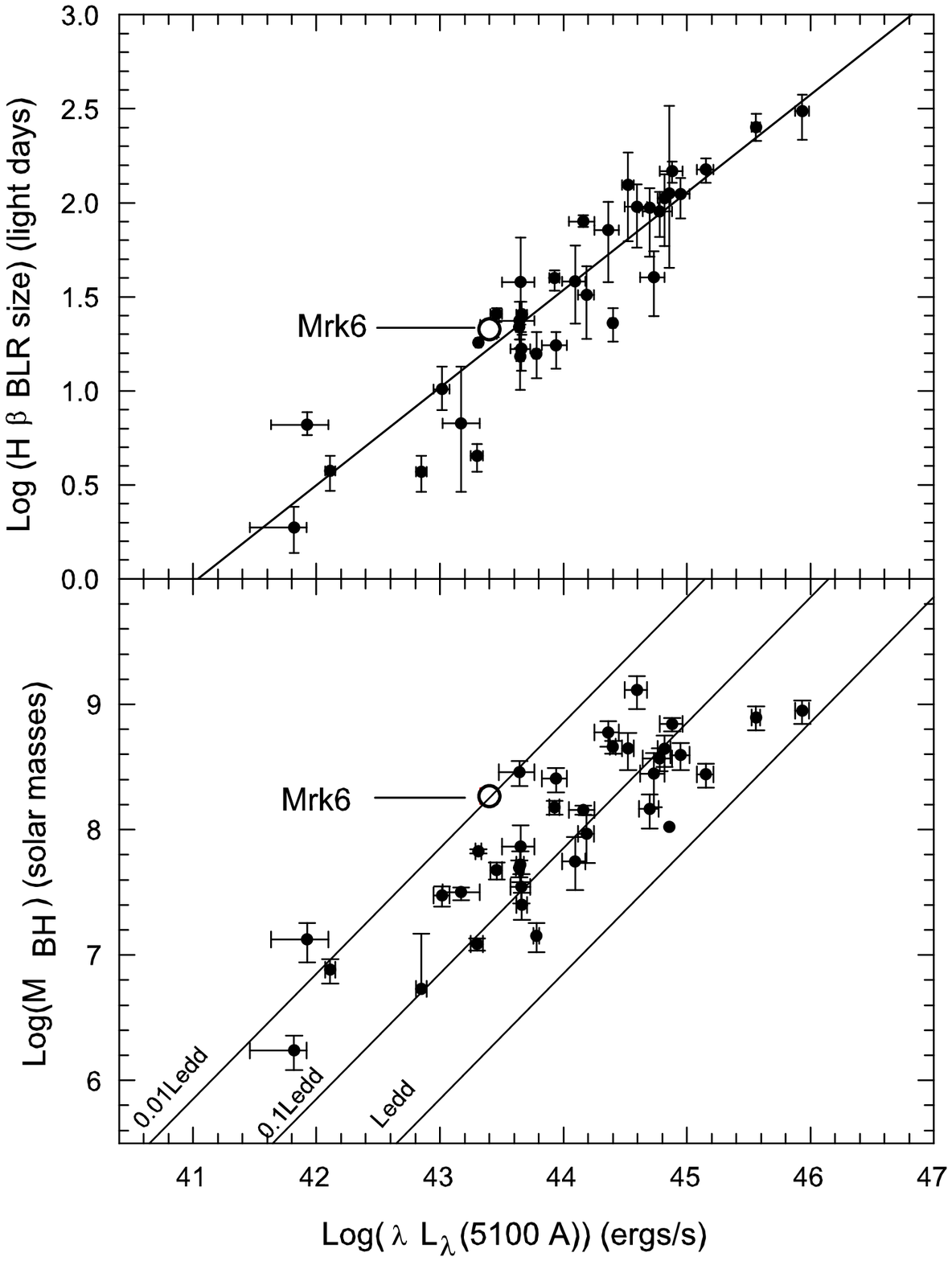}
\caption{Top: \hb\ BLR~size vs. luminosity at 5100\AA\
according to \citet{Bentz09a} and \citet{Denney10}. The luminosity of all nuclei was corrected for host galaxy contribution. The solid line is the best fit to the relationship log(R$_{BLR})$=$-$21.3+0.519 log(L). Bottom: the Mass--Luminosity diagram. The black hole mass of the majority of AGNs was taken from \citet{Peterson04}, except for Mrk~290, Mrk~817, NGC~3227, NGC~3516, NGC~4051, and NGC~5548, for which new results by \citet{Denney10} were used. Solid lines show the Eddington limit $L_{Edd}$ and its 10\% and 1\% fractions. The position of Mrk~6 nucleus is indicated on both plots.}
\label{sml}
\end{figure}
The bolometric luminosity of the Mrk~6 nucleus was adopted to be $L_{bol}\simeq 9\lambda L_{\lambda}(5100\AA)$ according to \cite{Kaspi2000} and it is equal to $L_{bol}(nucl)$=2.26
$\times10^{44}$erg\,s$^{-1}$. This luminosity is far from the Eddington limit (L$_{Edd}$), which
is equal to $L_{Edd}$=2.16$\times10^{46}$ erg\,s$^{-1}$ for a black hole mass of
1.8$\times10^8$ M$_{\odot}$. In other words, the Eddington ratio for Mrk~6 is ${L_{bol}/L_{Edd}\simeq0.01}$. In this case and because there are no clear indications of gas outflow from the BLR, the radiation pressure has a negligible effect on the reverberation mass estimate.

The position of the Mrk~6 nucleus on the BLR~Size--Luminosity diagram is shown in Fig.~\ref{sml}.
The BLR size and the luminosity of other galaxies in Fig.~\ref{sml} are taken from \citet{Bentz09a} and \citet{Denney10}. The black hole masses in Fig.~\ref{sml} are taken from~\cite{Peterson04}, except for the galaxies Mrk~290, Mrk~817, NGC~3227, NGC~3516, NGC~4051, for which we used new data from \citet{Denney10}.

\section{Velocity-resolved reverberation lags}
\label{vrrlag}
\subsection{Entire time interval: 1993--2008}
\label{vel-res lag} 
The question about whether the direction of gas motion can be determined from the response of the line profile to the continuum changes was firstly raised by \citet{Fabrika1980}. Generally speaking, the BLR gas velocity field can be random circular orbits, radial gas outflow or infall, or Keplerian motion. Examples demonstrating how the velocity resolved responses can be related to different types of BLR gas kinematics are given in \citet{Peterson01} and \citet{Bentz09b}. The random circular orbits generate a symmetric lag profile with the highest lag observed around zero velocity. The infall kinematics produces longer lags in the blue-shifted emission, and the outflow gas produces longer lags in the red-shifted emission.

Horne et al. (2004) formulated some important observational requirements for determining a
reliable velocity field of the BLR: (1) the time duration of observations should be at least
three times larger than the longest timescale of response, (2) the mean time between subsequent
observations should be at least two times less than the BLR light-crossing time, and (3) the
velocity sampling $\Delta V$ used for the  velocity-delay maps should be no less than the
spectral resolution of the data. According to \citet{Horne04}, such conditions can allow one to  distinguish clearly between alternative kinematic models of the BLR gas motion.
\begin{figure}
\includegraphics[width=84mm]{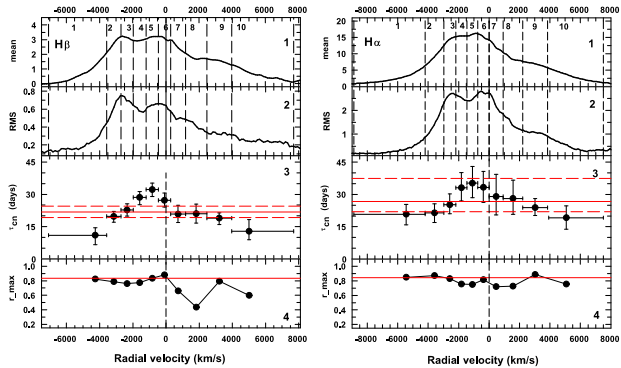}
\caption{Two top panels show the \hb\ mean and rms profiles (left) and the \ha\
mean profile (right) divided into ten bins of equal flux separated by vertical dashed lines. The flux units for the mean and rms profiles are 10$^{-13}$~ergs~cm$^{-2}$~s$^{-1}$~\AA$^{-1}$. The third panels from the top show the corresponding velocity-resolved time lag response, where the delays are plotted at the flux centroid of each velocity bin. Here the vertical error bars are 1$\sigma$ uncertainties in the lag for each velocity bin denoted by the horizontal error bars. The horizontal dotted and dashed lines on the third panels mark the mean centroid lag and 1$\sigma$ uncertainty, respectively, for the entire emission line. The bottom panels show the peak value of the correlation coefficient between the bin flux in the line and continuum. Here the horizontal dotted lines mark the correlation coefficient $r_{max}$ between the entire emission line and the continuum, as calculated in Table~\ref{ccf-ab}.}
\label{hablag-vel}
\end{figure}

In order to obtain the velocity-delay map, we measured the lag as a function of velocity in several bins across the line profile. We divided both the \ha\ and \hb\ lines into ten bins of equal flux, and the width of these bins was no less than 1000--1500 km s$^{-1}$. For each bin we calculated light curves from the Balmer line fluxes. Then each of these light curves was cross-correlated with the continuum light curve following the same procedure as described in Section~\ref{fccf}.  Figure~\ref{hablag-vel} shows hydrogen line profiles (mean and rms) subdivided into bins (two upper panels). The two middle panels demonstrate the lag measurements for each of the bins. The vertical error bars show $1\sigma$ uncertainties for the time lag, and the horizontal bars represent the bin width. The horizontal solid and dashed lines in the two middle panels show the mean BLR lag and associated errors as listed in Table~\ref{ccf-ab}. The bottom panels show the peak correlation coefficient between the bin flux in the line and continuum, $r_{max}$. Figure~\ref{hablag-vel} shows that
\begin{enumerate}
\renewcommand{\theenumi}{(\arabic{enumi})}
\item The mean and rms profiles of \hb\, and \ha\, are not symmetric with respect to zero
velocity. The centroid of the mean and rms profiles is shifted to the short-wave part of the line. The variable parts of \hb\ and \ha\ have two well-defined peaks, one of them is almost central (between 5 and 6 bins) and another is blue-shifted. In addition, there is a weaker peak in the red part of the line profile.

\item The time delay between the higher velocity gas in the BLR and the continuum is shorter
than the delay between the low velocity gas and the continuum. Such a behaviour is typical for
virialized gas motions.

\item The lag in the blue wing of the \hb\ line is greater than the lag in the red side of this line. The \ha\ velocity-resolved lags shows the same tendency. This is consistent with expectations from the infall model of gas motion. Thus, it is possible we have virialized motion combined with infall signatures.

\item The correlation coefficient of different segments of the lines is different. The bin corresponding to a radial velocity of $V_r \approx +1500\,km\,s^{-1}$ shows poor correlation with the
continuum variation, especially in \hb. This fact was earlier noted by \citet{Ser99}.
\end{enumerate}

\begin{figure*}
\includegraphics[width=170mm]{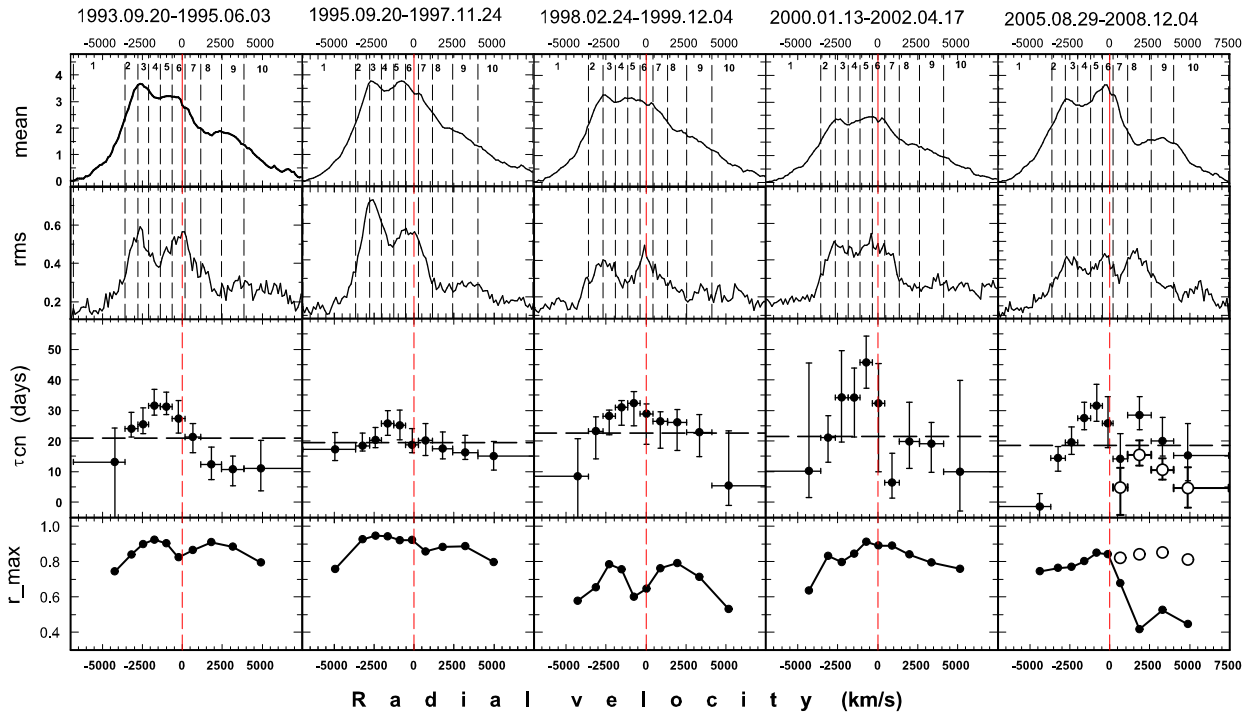}
\caption{The same as in Figure~\ref{hablag-vel}, but for the \hb\ line for the five
time intervals.}
\label{hblag-5p}
\end{figure*}

\subsection{Velocity delay maps in the five time intervals}
We have computed the velocity-resolved time delays for the five time intervals given in Table \ref{time intervals}. For each subset we made the velocity-dependent cross-correlation analysis for the \hb\ line profile bins as described in the previous section. The \hb\ line was selected because its sampling is better. In Figure~\ref{hblag-5p} the mean and rms \hb\ profiles, the velocity-resolved time lag response, and the velocity-dependent peak correlation coefficient are shown for the five time intervals. Upon inspection of Figure~\ref{hblag-5p} it becomes clear that the mean and rms profiles are different among the five periods. The relative intensity in the blue peak and in the central peak changes very strongly: during the first interval the blue peak is higher then the red one. The opposite situation is seen in the fifth interval. The flux in continuum as well the flux in the \hb\ line systematically decreased from the second to the fourth time interval, as is seen in Figure~\ref{lc} and Figure~\ref{ccf-b}. The \hb\ rms profile shows two peaks in the first, second and third period, the flat top in the fourth period, and in the fifth period we see three peaks.

Figure~\ref{hablag-vel} demonstrates that the high-velocity gas in the wings exhibits a shorter lag than the low-velocity gas, supporting the virial nature of gas motion in BLR: the gas kinematics that is dominated by the central massive object.

\begin{figure}
\includegraphics[width=84mm]{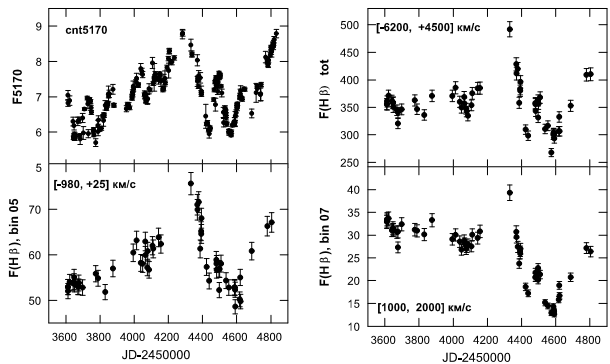}
\caption{Light curves in the continuum and \hb\, in 2005--2008 over the velocity range from $-6200$ to $+4500~km\,s^{-1}$ (top panels) as well as the light curves of the broad \hb\ line in bin~5 in the velocity range from $-980$ to $+25~km\,s^{-1}$, and in the bin~7 in the velocity range from 1000 to 2000~km\,s$^{-1}$ (bottom panels). The continuum fluxes are in units of $10^{-15}$~erg\,s$^{-1}$\,cm$^{-2}$\,\AA$^{-1}$, and the broad emission line fluxes are in units of $10^{-15}$~erg\,s$^{-1}$\,cm$^{-2}$.}
\label{hblc05-08}
\end{figure}

However, the lag is slightly larger in the blue wing than in the red wing for all subintervals.  
This is a signature of the infall gas motion. In the fifth period (2005--2008) the velocity delay map is more symmetric, but the seventh bin shows very small lag, as well as in the previous time interval. For 2005--2008 there is a poor correlation with the continuum for bins~7--10. A more detailed examination of the bin light curves for 2005--2008 (Figure~\ref{hblc05-08}) revealed that there is a trend for the \hb\ flux in bins 7--10, which is almost absent in the bins~1--6.  Following the advice of our reviewer we removed the trend from the \hb\ light curves for bins 7--10. No more significant trends were found for other time intervals. 
In Figure~\ref{hblag-5p} the detrended lags and correlation coefficients are shown by open circles. After detrending procedure, the lag--velocity dependence became more similar to the lag--velocity dependence for the first period, for which the difference in lag between the blue and red wings is largest.

So, it is most likely that the BLR kinematic in Mrk~6 is a combination of the Keplerian gas motion and infall gas motion.

\section{Conclusion}
\label{sum}
We have reported our new results on the Mrk~6 nucleus from 1998--2008 observational data together with the previous results published by \citet{Ser99}. We found that
\begin{enumerate}
 \renewcommand{\theenumi}{(\arabic{enumi})}
 
\item The flux of the Mrk~6 nucleus in 1992--2008 varied significantly in the continuum as well as in the \ha\ and \hb\ broad emission lines. The relative amplitude of the continuum flux variability is larger than in the hydrogen lines, and it is greater in the \hb\ line than in the \ha\ (see Table~\ref{stat}). It is typical for the most of Seyfert galaxies. This agrees with the predictions of \citet{Korista04} based on new photo-ionization calculations of the BLR-like gas. 

\item We found the average time delay between the total \hb\ flux and the continuum flux at 5170~\AA\ to be $21.1 \pm 1.8$ days, and the time delay does not vary significant among individual time intervals. It seems that the size of the \hb\ emission region remains approximately the same over long time periods. The \ha\ flux responds to the changes in the $F5170$ continuum with a lag of $26.8^{+10.6}_{-4.8}$ days. 

\item When the continuum flux varies, the photo-ionization models predict the existence of the relation between the BLR size and the luminosity. However, because the large uncertainties in the lag for individual time intervals we are unable to find such a relation. For the same reason, it is unable to obtain a dependence between the lag and line width, and it is impossible to check whether this dependence is consistent with the dependence $V\propto r^{-1/2}$ expected for the gravitationally dominated motion.

\item The \hb\ line width is larger than that of \ha. This is naturally explained by photo-ionization calculations \citep[e.g.,][]{Korista04}: the effective emission region of \hb\ is smaller and closer to the ionizing source than the effective region of \ha, and the gas velocities in \hb\ are higher.

\item By examining the velocity-resolved lags for the broad \hb\ and \ha\ lines, we found that the lag in the high-velocity wings are shorter than in the line core. This indicates virial motions of gas in the BLR. However, the lag is slightly larger in the blue wing than in the red wing for the entire time interval as well as for the individual periods considered in the present paper. This is a signature of the infall gas motion. Probably the BLR kinematic in the Mrk~6 nucleus is a combination of the Keplerian gas motion and infall gas motion.

\item Some profile segments often show poor correlation with the continuum flux. According to \citet{Gaskell10} this effect can arise because off-axis sources of ionizing continuum flux can appear, which might not make a detectable contribution to the total continuum flux variability, but they will have an influence on the line only over a narrow range of radial velocity in the BLR. If these local off-axis events will vary out of phase with the variability of the dominant source, the result will be to give a weak correlation between the continuum flux and the line flux in the narrow range of radial velocity. 

\item We determined the black hole mass from the lag and line width measurements of the \hb\ and \ha\ lines. The mass was found to be $M_{BH} = (1.8\pm 0.2)\times 10^8\,M_{\odot}$ for the \hb\ line and slightly greater and less reliable from the \ha\ line. Under such a mass and the luminosity of $\lambda L_{\lambda}$(5100)=(2.51$\pm$0.38)$\times10^{43}$~erg\,s$^{-1}$, the Mrk~6 nucleus is located on the upper edge of the Mass-Luminosity diagram that corresponds to the Eddington ratio of about 0.01. This confirms the assumption  \citep[e.g.,][]{Ser11} that there is anticorrelation between broad-line widths and Eddington luminosity ratio $L_{bol}/L_{Edd}$. The Mrk~6  position on the BLR~Size--Luminosity diagram does not contradict the fit $R_{BLR}\propto L^{0.5}$ determined by \citet{Bentz09a}.
\end{enumerate}

\section*{Acknowledgments}
We thank the anonymous reviewer for useful comments and suggestions. We also thank S. Nazarov and the staff of 2.6-m and 0.7-m telescope for help during our observations. SSG acknowledges the support to CrAO in the frame of the `CosmoMicroPhysics'  Target Scientific Research Complex Programme of the National Academy of Sciences of Ukraine (2007--2012). VTD acknowledges the support of the Russian Foundation of Research (RFBR, project no.~09-02-01136a). The CrAO CCD cameras were purchased through the US Civilian Research and Development for Independent States of the Former Soviet Union (CRDF) awards UP1-2116 and UP1-2549-CR-03.

\label{lastpage}
\end{document}